\documentclass[article]{IEEEtran}
\IEEEoverridecommandlockouts
\usepackage{cite}
\usepackage{amsmath,amssymb,amsfonts}
\usepackage{algorithmic}
\usepackage{graphicx}
\usepackage{textcomp}
\usepackage{xcolor}

\usepackage{float}
\usepackage{color}
\usepackage{amssymb} 
\usepackage{verbatim}

\usepackage{caption}
\usepackage{subcaption}
\usepackage{float}
\usepackage{gensymb} 

\newtheorem{theorem}{ Theorem}
\newtheorem{lemma}{ Lemma}

\usepackage{multirow}

\usepackage{tabularx} 
\usepackage{array}     
\usepackage{makecell}  
\newcolumntype{Y}{>{\centering\arraybackslash}X}  

\usepackage{stfloats}
\def\BibTeX{{\rm B\kern-.05em{\sc i\kern-.025em b}\kern-.08em
		T\kern-.1667em\lower.7ex\hbox{E}\kern-.125emX}}

\begin{document}

\title{
	Integrated Localization and Communication with Sparse MIMO: Will Virtual Array Technology also Benefit Wireless Communication? \\
	
		
		\thanks{
			Part of this work has been presented at the 2024 IEEE WCSP, Hefei, China, in Oct. 2024 \cite{10404673}.			
			
			Hongqi Min, Xinrui Li and Yong Zeng are with the National Mobile Communications Research Laboratory, Southeast University, Nanjing 210096, China. Yong Zeng is also with the Purple Mountain Laboratories, Nanjing 211111, China (E-mail: {minhq, xinrui\_li, yong\_zeng}@seu.edu.cn). (Corresponding author: Yong Zeng.)
			
			Ruoguang Li is with the College of Information Science and Engineering,
			Hohai University, Changzhou 213200, China (E-mail: ruoguangli@
			hhu.edu.cn).
			}
}

\author{Hongqi Min, Xinrui Li, Ruoguang Li, \emph{Member, IEEE} and Yong Zeng, 
\emph{Fellow, IEEE}}

\maketitle

\begin{abstract}
	For the sixth generation (6G) wireless networks, achieving high-performance integrated localization and communication (ILAC) is critical to unlock the full potential of wireless networks. To simultaneously enhance wireless localization and communication performance cost-effectively, this paper proposes sparse multiple-input multiple-output (MIMO) based ILAC with nested and co-prime sparse arrays deployed at the base station (BS). Sparse MIMO relaxes the traditional half-wavelength antenna spacing constraint to enlarge the antenna aperture, thus enhancing localization degrees of freedom (DoFs) and providing finer spatial resolution. However, it also leads to undesired grating lobes, which may cause severe inter-user interference (IUI) for communication and angular ambiguity for localization.
	While the latter issue can be effectively addressed by the virtual array technology, by forming sum or difference co-arrays via signal (conjugate) correlation among array elements, it is unclear whether the similar virtual array technology also benefits wireless communications for ILAC systems. 
	In this paper, we first reveal that the answer to the above question is negative, by showing that forming virtual arrays for wireless communication will cause destruction of phase information, degradation of signal-to-noise ratio and aggravation of multi-user interference. 
	Therefore, we propose the so-called hybrid processing for sparse MIMO based ILAC, i.e., physical array based communication while virtual array based localization. 
	To this end, we characterize the beam pattern of sparse arrays by three metrics, i.e., main lobe beam width, peak-to-local-minimum ratio, and side lobe height, demonstrating that despite of the introduction of grating lobes, sparse arrays can also bring benefits to communications thanks to its narrower main lobe beam width than the conventional compact arrays.	
	Extensive simulation results are presented to demonstrate the performance gains of sparse MIMO based ILAC over that based on the conventional compact MIMO.
\end{abstract}


\section{Introduction}
As the standardization of the sixth generation (6G) mobile communication networks approaching, one of the 6G typical usage scenarios defined by the international telecommunication union (ITU)\textemdash integrated sensing and communication (ISAC)\textemdash has garnered significant research attention \cite{9737357}. 
Meanwhile, unlike ISAC for which the targets to be sensed only passively reflect/scatter the signal without active involvement, for integrated localization and communication (ILAC) \cite{zhiqiangxiao1}, the targets to be localized actively participate the process by transmitting/receiving the wireless signals. In fact, ILAC has been around for mobile communication networks since the second generation (2G). With the growing demand for hyper-reliable and low-latency communications, the need for high-precision localization has further increased, making ILAC critical to unlock the full potential of wireless networks \cite{zhiqiangxiao1}.
In ILAC, wireless localization and communication functions will be deeply integrated, sharing the same network infrastructure, hardware platform, wireless resources, and even the protocol. 
This helps to not only reduce costs, enhance both communication and localization performance, but also support a broader range of applications in future wireless networks.
As one of the potential technologies for 6G, extremely large-scale MIMO (XL-MIMO) is a promising candidate that can boost sensing, localization and communication performance by significantly increasing the number of antennas \cite{10496996,9903389,10545312}.
It is able to provide high resolution in the spatial domain, enlarge the near-field region and enhance the spatial multiplexing gain, beamforming gain, diversity gain\textemdash ultimately improving both communication, sensing and localization capabilities \cite{wangEnhancingSpatialMultiplexing2024a}.

However, if XL-MIMO is implemented as the conventional compact antenna arrangement, i.e., the inter-antenna spacing is restricted to half-wavelength, it would incur prohibitive hardware cost, power consumption, and signal processing overhead. 
Note that using half-wavelength antenna spacing for MIMO communications is due to three main reasons\cite{xinruiLiSparseMIMO}. 
Firstly, it ensures a sufficient distance between adjacent elements to avoid severe mutual coupling. Secondly, in rich scattering scenarios, half-wavelength is approximately the channel coherence distance, allowing for optimal spatial diversity gain. Thirdly, half-wavelength represents the maximum spacing that prevents the occurrence of grating lobes in the angular response of the array.
On the other hand, sparse MIMO removes the half-wavelength spacing restriction, enabling the array aperture to be enlarged without increasing the number of antenna elements or radio frequency (RF) chains \cite{xinruiLiSparseMIMO}. 
As revealed in \cite{xinruiLiSparseMIMO}, sparse MIMO for ISAC/ILAC offers several appealing advantages, including finer spatial resolution, larger sensing degrees of freedom (DoFs), enlarged near-field region, reduced mutual coupling, more flexible deployment, as well as savings of hardware, energy, and signal processing costs. 
Note that for MIMO wireless localization or sensing, DoF represents the number of independent targets or sensing parameters that can be estimated simultaneously, which is different from the concept of spatial multiplexing gain in MIMO wireless communications.

However, for sparse MIMO wireless systems, directly increasing the antenna spacing larger than half-wavelength results in sparse sampling in the spatial domain, which forms undesirable grating lobes in the beam pattern. By viewing spatial sampling as the analogy to the classic temporal sampling, it can be easily inferred that sparse spatial sampling with sampling interval violating Nyquist condition leads to aliasing, here manifested as the appearance of grating lobes within the $\left[-\pi/2, \pi/2\right]$ in angular domain \cite{johnson1993array,9947035}. 
For localization or sensing, these grating lobes cause angular ambiguity in algorithms based on angular power spectrum, such as fast Fourier transform (FFT), Bartlett, and multiple signal classification (MUSIC), etc \cite{7815358, 1017528,1143830}. For wireless communications, on the other hand, grating lobes also lead to severe inter-user interference (IUI) \cite{10545312}.

Fortunately, for sparse MIMO based wireless sensing or localization, it is well-known that the so-called virtual array technology can effectively address the angular ambiguity issue. For example, by utilizing correlation or conjugate correlation for signals from different sparse array elements, sum or difference co-arrays can be created, whose virtual array element positions depend mainly on the sums or differences of the physical array element positions. 
These co-arrays also refer to virtual arrays, providing larger virtual apertures compared to physical apertures, thereby enhancing the resolution and DoFs in sensing and localization. 
For wireless localization or sensing, sparse MIMO with various array architectures has been studied, which can be generally classified into: uniform sparse array (USA) and non-uniform sparse array (NUSA) \cite{xinruiLiSparseMIMO}, while typical examples of the latter include nested array (NA)\cite{5456168}, co-prime array (CPA) \cite{vaidyanathanSparseSensingCoPrime2011b}, minimum redundancy array (MRA)\cite{1139138}, modular array (MoA)\cite{9665444}, etc. 
Additionally, there are sparse arrays where antenna positions are optimized based on metrics such as main lobe beam width, side lobe levels and null steering positions \cite{huanSASASuperResolutionAmbiguityFree2023}.
While USA and MoA are designed mainly for simplicity of implementation, without considering to form continuous virtual array for sensing or localization \cite{xinruiLiSparseMIMO}, MRA gives largest virtual array by minimizing redundant information between array elements. However, the element positions of MRA can only be obtained through numerical search, whose complexity increases sharply as the number of antennas $M$ grows. 
Besides, geometry optimization-based sparse MIMO requires high optimization complexity. 
On the other hand, sparse MIMO based on nested and co-prime arrays  offers strong localization and sensing performance in terms of DoFs and resolution. In particular, for the two-level nested array, an $N_1$-element compact uniform linear array (ULA) with antenna elements separated by $d_0=\lambda/2$ and an $N_2$-element sparse ULA with elements separated by $\eta d_0$ are concatenated, where $\lambda$ is the wavelength and $\eta=N_1+1$ denotes the sparsity of the sparse ULA \cite{5456168}.
On the other hand, the co-prime array consists of a $2M_1$-element sparse ULA with sparsity $\eta_1=M_2$ and an $M_2$-element sparse ULA with sparsity $\eta_2=M_1$.
Both nested and co-prime arrays can provide $\mathcal{O}\left(M^2\right)$ localization DoFs using only $\mathcal{O}\left(M\right)$ physical array elements \cite{5456168}. 

On the other hand, for wireless communications, to the authors' best knowledge, it is unclear whether the similar virtual array technology, which works well for sensing and localization, can also be exploited to address the undesired grating lobes and benefit wireless communications. Note that this question is becoming rather relevant for ILAC systems for which the same set of antenna array needs to be shared between localization and communication. In this paper, we reveal that the answer to the above question is negative, and the main reasons are three folds. 
Firstly, creating virtual arrays requires signal correlation or conjugate correlation across different array elements, which causes the loss of phase information of the information-bearing symbols. While this is fine for wireless localization or sensing since they mainly concern the target parameters, such as the target angle, Doppler, and distance, it is detrimental for the information-bearing capability for communications.
Secondly, creating virtual arrays also leads to correlation between the useful signal and effective noise, leading to a degradation in the signal-to-noise ratio (SNR) compared to the SNR received by the physical array. 
Thirdly, in multi-user scenarios, the number of interference and noise terms in the virtual array domain increases, where interference can be split into two types of IUI. It is difficult for the virtual array to effectively suppress both types of interference simultaneously.
Fortunately, despite the presence of grating lobes, it has been shown in \cite{wangEnhancingSpatialMultiplexing2024a, 10465094} that a USA can achieve better communication performance than its compact counterpart, particularly in hot-spot areas with densely located users. 
This is because sparse array provides higher spatial resolution through narrower main lobe beam width, and the probability of users being located at higher-order grating lobes is relatively low, allowing the resulting interference to be effectively suppressed.	
Our previous work in \cite{10404673} also demonstrated that sparse MIMO based on nested arrays can achieve higher communication performance, while preserving their sensing advantages compared with conventional compact MIMO.
Sparse arrays have also been investigated in near-field communication \cite{10545312,zhou2024sparse}, where they illuminate improved beam focusing characteristics in the main lobe, effectively suppressing interference and enhancing communication performance.
Additionally, sparse MIMO is able to enhance the near-field effective DoF, as demonstrated in \cite{wangEnhancingSpatialMultiplexing2024a}.
The main contributions of this paper are summarized as follows:
\begin{itemize}
	\item First, we investigate nested and co-prime array based sparse MIMO for ILAC systems, where an ILAC base station (BS) equipped with sparse array concurrently supports multiple communication and localization users. The question whether virtual array technology can also benefit communication is addressed and we show that the answer is negative, which is demonstrated from three aspects: destruction of phase information, degradation of SNR and aggravation of multi-user interference. 
	Therefore, we propose the so-called hybrid processing for sparse MIMO based ILAC, i.e., physical array based communication while virtual array based localization. 
	
	\item Next, we analyze the communication performance of sparse MIMO, including nested and co-prime arrays, by characterizing the beam pattern of the physical array. 
	As the beam pattern of sparse array differs from that of traditional compact ULA, where the latter has evenly spaced null steering points and smoothly descending side lobes, we define three new beam pattern metrics for sparse array: main lobe beam width ($\mathrm{BW}$), peak-to-local-minimum ratio ($\mathrm{PLMR}$), and side lobe height ($\mathrm{SLH}$). 
	Then, we derive the bounds of these metrics for different sparse array architectures, which depend on sparse array configurations $\left(N_1, N_2\right)$ and $\left(M_1, M_2\right)$.
	It is revealed that sparse arrays can bring benefits to communications even though there exists grating lobes thanks to the narrower main lobe beam width being $\frac{1}{\mathcal{O}\left(M^2\right)}$, compared to $\frac{1}{\mathcal{O}\left(M\right)}$ of compact MIMO.	
	However, such a benefit could be compromised by its lower $\mathrm{PLMR}$ and dominant grating lobes.
	
	\item Lastly, we simulate the communication and localization performance of sparse MIMO with an ILAC system, demonstrating the great potential of sparse MIMO for ILAC.
	We first demonstrate that with the same number of array elements, different sparse array configurations yield varying communication performance, as both the physical and virtual array architectures are different.	
	For nested arrays, the larger outer subarray may result in a more noticeable performance improvement, while the larger subarray 1 is better for co-prime arrays.
	Then, the joint communication and localization performance of the sparse arrays are presented and a set of sparse array configurations satisfying Pareto condition is provided.
	On the other hand, for different user distribution scenarios, the results show that sparse arrays can always achieve better localization resolution and DoF compared to conventional compact ULAs. 
	The communication performance is notably enhanced for scenarios with densely located users due to the higher spatial resolution of sparse arrays. However, only when the user distribution is highly sparse may sparse arrays experience a slight loss in communication performance due to grating lobe issues compared to their compact counterparts.
\end{itemize}

\emph{Notations:} Vectors and matrices are denoted by boldface lower- and upper-case letters, respectively. $\|\cdot\|$ denotes the Frobenius norm, while $(\cdot)^T$ and $(\cdot)^H$ are the matrix transpose and Hermitian transpose operators, respectively. $\mathbb{C}$, $\mathcal{N}$, $\mathcal{CN}$ denote the set of complex number, the Gaussian distribution and the circularly symmetric complex Gaussian distribution, respectively. $\mathbf{I}_n$ represents the $n\times n$ identity matrix. $\otimes$ and $\odot$ represent the Kronecker product and Khatri-Rao (KR) product, respectively. $E\left\{\cdot\right\}$ and $var\left\{\cdot\right\}$ denote the expectation and variance operator, respectively.

\section{System Model} \label{wuyifan}
\subsection{System Model}
\begin{figure}[h]
	\centering
	\includegraphics[width = 7cm]{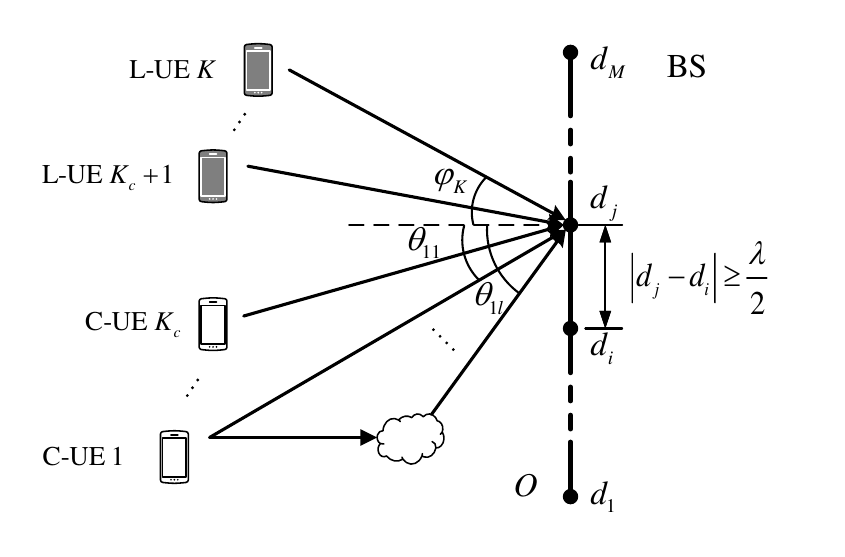}
	\caption{Uplink ILAC system with sparse array deployed at BS.}
	\label{system_1}
\end{figure}
As illustrated in Fig. \ref{system_1}, we consider an uplink ILAC system consisting of an ILAC BS and $K$ user equipments (UEs), $K_c$ of which are communication UE (C-UE) and the rest are localization UE (L-UE). The BS is equipped with an $M$-element sparse array, which needs to concurrently support the uplink communication for the C-UEs and estimate angle-of-arrival (AoA) for all the L-UEs.
A reference point $O$ is set at the origin of the linear grid. Define the position of the $m$-th antenna as $d_m, m=1,2,\ldots,M$ and the first antenna of the sparse array is assumed at the reference point $\mathcal{O}$, thus $d_1=0$. 
Note that since sparse MIMO removes the restriction of half-wavelength antenna spacing, making the adjacent antennas can be separated by larger than half-wavelength, thus $|d_i-d_j|\geq d_0, i\neq j$.
For convenience, the antenna locations of a sparse array can be expressed as a position index set $\mathcal{D}=\left\{\bar{d}_1, \bar{d}_2, \ldots, \bar{d}_M\right\}$, where $\bar{d}_m=d_m / d_0$.
Under the far-field assumption, the array steering vector is
\begin{equation}\label{chenyiyi}
	\mathbf{a}\left(\theta\right)=\left[1,e^{j\pi\bar{d}_2\sin\theta},\ldots,e^{j\pi\bar{d}_M\sin\theta}\right]^{T},
\end{equation}
where $\theta$ is the AoA of the signal.

The channel vector between C-UE $k$ and the BS can be expressed as 
\begin{equation}\label{minhongguang}
	\mathbf{h}_k=\sum\nolimits_{l=1}^{L_k}\beta_{kl}\mathbf{a}(\theta_{kl}) =\mathbf{A}_{ck}\boldsymbol{\beta}_{ck},\quad k=1,2,\ldots, K_c, 
\end{equation} 
where  $\mathbf{A}_{ck}=\left[ \mathbf{a}(\theta_{k1}),\ldots,\mathbf{a}(\theta_{kL_k})\right] $ and $\boldsymbol{\beta}_{ck} = \left[\beta_{k1},\ldots,\beta_{kL_k}\right]$. Besides, $\mathbf{a}\left(\theta_{k l}\right)$ and $\beta_{kl}$ are the array steering vector and the complex-valued channel gain of the $l$-th path of C-UE $k$. $L_k$ is the number of multi-paths of C-UE $k$ and $\theta_{kl}$ represents its AoA of the $l$-th path.

For the links of L-UEs, it is assumed that each L-UE has line-of-sight (LoS) component, so that the channel between the $q$th L-UE and the BS can be expressed as 
\begin{equation}
	\mathbf{h}_q=\beta_q\mathbf{a}\left(\varphi_{q}\right),\quad q=K_c+1,\ldots, K,
\end{equation}
where $\beta_q$ and $\varphi_{q}$ represent the channel gain and AoA of L-UE $q$, respectively. 	

The received signal at the BS can be expressed as
\begin{equation}
	\label{zhangguangjun}
	\mathbf{y}\left[n\right]=\sum\nolimits_{k=1}^{K} \mathbf{h}_k \sqrt{P_k} x_k\left[n\right] +\mathbf{n}\left[n\right],
\end{equation}
where $x_k[n], 1\leq k\leq K_c$, denotes the independent and identically distributed (i.i.d.) information-bearing symbol of C-UE $k$, and  $x_k[n], K_c+1\leq k\leq K$, represents localization signal of L-UE $k$ with $E\{|x_k[n]|^2\}=1$;
$P_k$ denotes the transmit power of UE $k$;
$\mathbf{n}[n] \sim \mathcal{C} \mathcal{N}(0, \sigma^2 \mathbf{I}_M)$ is the additive white Gaussian noise (AWGN). 

\subsection{Sparse Array Architecture}
\begin{figure}[htbp]
	\centering
	\begin{subfigure}{1\linewidth}
		\centering
		\includegraphics[width=0.9\linewidth]{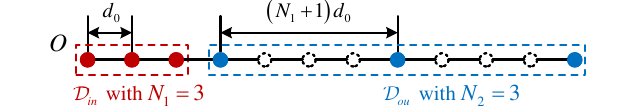}
		\caption{Nested array with $\left(N_1, N_2\right)=\left(3, 3\right)$.}
		\label{array_structure_nested}
	\end{subfigure}
	\centering
	\begin{subfigure}{1\linewidth}
		\centering
		\includegraphics[width=0.9\linewidth]{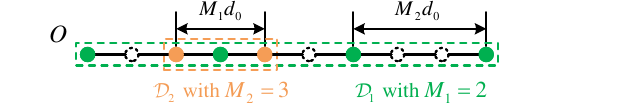}
		\caption{Co-prime array with $\left(M_1, M_2\right)=\left(2, 3\right)$.}
		\label{array_structure_coprime}
	\end{subfigure}
	\caption{Illustration of nested and co-prime arrays with $M=6$ physical antenna elements.}
	\label{lixiang}
\end{figure}

We consider nested and co-prime sparse arrays, as illustrated in Fig. \ref{lixiang}. 
An $M$-element nested array consists of an $N_1$-element inner subarray with inter-element spacing $d_0$ and an $N_2$-element outer subarray with inter-element spacing $\left(N_1+1\right)d_0$, where $M=N_1+N_2$ \cite{5456168}.

\emph{Two-level Nested Array}: A nested array is defined in terms of the non-negative integer parameter pair $\left(N_1, N_2\right)$, which is the union of a compact ULA and a sparse ULA, i.e., $\mathcal{D}_{\mathrm{na}}=\mathcal{D}_{in}\cup \mathcal{D}_{ou}$:
\begin{equation}
	\resizebox{0.9\hsize}{!}{$
		\begin{aligned}
			\text{Compact ULA } \mathcal{D}_{in}:& \left\{0,1, \ldots, N_1-1\right\},\\
			\text{Sparse ULA } \mathcal{D}_{ou}:&\left\{N_1, 2 N_1+1, \ldots, N_1N_2+N_2-1\right\}.
		\end{aligned}$}
\end{equation}
Note that the nested array can degenerate to the traditional compact ULA when $(N_1, N_2)=(0,M), (M-1,1)\text{ or }(M,0)$.

For co-prime array, it consists of a $2M_1$-element sparse ULA subarray with inter-element spacing $M_2d_0$ and an $M_2$-element sparse ULA subarray with inter-element spacing $M_1d_0$. Note that the first antenna of the two subarrays is shared, thus $M=2M_1+M_2-1$ \cite{vaidyanathanSparseSensingCoPrime2011b, palCoprimeSamplingMusic2011}.

\emph{Co-prime Array}: A co-prime array is defined in terms of the co-prime integer parameter pair $\left(M_1, M_2\right)$, and $M_1<M_2$ is assumed without loss of generality. It is the union of two sparse ULA subarrays, i.e., $\mathcal{D}_{\mathrm{cp}}=\mathcal{D}_{1}\cup \mathcal{D}_{2}$:
\begin{equation}
	\resizebox{0.8\hsize}{!}{$
		\begin{aligned}
			\text{Sparse ULA } \mathcal{D}_{1}:& \left\{0, M_2, 2M_2, \ldots, \left(2M_1-1\right)M_2\right\},\\
			\text{Sparse ULA } \mathcal{D}_{2}:&\left\{M_1, 2M_1, \ldots, \left(M_2-1\right)M_1\right\}.    
		\end{aligned}$}
\end{equation}
Note that the co-prime array degenerates to compact ULA when $M_1=1$. 

\subsection{Beam Pattern}
Based on the aforementioned array steering vector in (\ref{chenyiyi}), the beam pattern is defined as the normalized power pattern at an arbitrary observation direction $\theta$ with respect to the beamforming direction $\theta_0$ \cite{xinruiLiSparseMIMO}, i.e.,  
\begin{equation}\label{telangpu}
	\resizebox{0.91\hsize}{!}{$\begin{aligned}
			& G\left(\Delta_{ki};\mathcal{D}\right) = \frac{ \left|\mathbf{a}^H\left(\theta_k\right)  \mathbf{a}\left(\theta_i\right)\right|^2 }{\| \mathbf{a}\left(\theta_k\right)\|^2 \| \mathbf{a}\left(\theta_i\right)\|^2 } \\
			& = \frac{1}{M^2}\left|\sum\nolimits_{m=1}^{M}e^{j\pi \bar{d}_m\left(\sin \theta_k - \sin \theta_i\right)}\right|^2
			\triangleq \frac{1}{M^2}\left|\sum\nolimits_{m=1}^{M}e^{j\pi \bar{d}_m\Delta_{ki}}\right|^2,
	\end{aligned}$}
\end{equation}
where $\Delta_{ki}\triangleq\sin \theta_k - \sin \theta_i $ is defined as the spatial angle difference and $\mathcal{D}$ is the antenna position index set.
$G\left(\Delta_{ki};\mathcal{D}\right)$ has similar mathematical expression to the angular ambiguity function \cite{ericAmbiguityCharacterizationArbitrary1998} and power form of array factor \cite{balanis2015antenna}. 
These two definitions also represent the intrinsic information conveyed by the beam pattern. The former indicates the array's ability to distinguish targets from the $\theta_k$ and $\theta_i$ directions, where $\theta_i$ represents the true AoA and $\theta_k$ denotes the estimated AoA. This distinction is measured by the spatial correlation of the steering vectors, which plays a critical role in the performance of wireless sensing or localization systems. The latter represents the far-field strength in the $\theta_k$ direction when the beam is formed towards the $\theta_i$ direction for phased array systems, where $\theta_k$ and $\theta_i$ denote the desired and interfering signal directions, respectively. This strength is calculated by the square of the magnitude of the electric or magnetic field as a function of the angular direction.
 
\subsection{Virtual Array Based Localization Signal Processing}
In the field of wireless sensing or localization, sparse MIMO can achieve a virtual array through baseband signal processing \cite{5456168, vaidyanathanSparseSensingCoPrime2011b}.
Specifically, the second-order statistical data of the received signal can be obtained by calculating the (conjugate) correlation of the received signal, which yields an equivalent virtual array's received signal. If the elements of the physical array are arranged according to specific rules, such as nested or co-prime structures, a compact ULA virtual array can be formed, with a number of elements far greater than the original physical array's element count \cite{5456168}.
Therefore, localization signal processing can be performed on the virtual array domain.

It is assumed that the communication symbols have been detected based on communication signal processing, and we subtract the communication signals in ($\ref{zhangguangjun}$), leaving only the localization signals as
\begin{equation} \label{youxiaohu}
	\begin{aligned}
		\mathbf{y}_l\left[n\right]
		&= \sum\nolimits_{k=K_c+1}^{K}\mathbf{h}_k\sqrt{P_k}x_k\left[n\right] + \mathbf{n}\left[n\right]\\
		&=\mathbf{A}_l\mathbf{s}_l\left[n\right]+\mathbf{n}\left[n\right],
	\end{aligned}
\end{equation}
where  $\mathbf{s}_l\left[n\right]=\left[\beta_{K_c+1}\sqrt{P_{K_c+1}} x_{K_c+1}\left[n\right], \ldots,\beta_{K}\sqrt{P_{K}} \right.$ $\left.x_{K}\left[n\right]\right]^T\in \mathbb{C}^{\left(K-K_c\right)\times1}$ and $\mathbf{A}_l = \left[ \mathbf{a}(\theta_{K_c+1}),\ldots,\mathbf{a}(\theta_{K})\right] \in \mathbb{C}^{M\times\left(K-K_c\right)}$.
Note that classic AoA estimation algorithms cannot be directly performed on (\ref{youxiaohu}) due to the grating lobes introduced by the non-uniform sparse spatial sampling. Specifically, the beamspace based algorithms requiring spectrum peak search, such as FFT, Bartlett, and MUSIC, will suffer from angular ambiguity. Besides, the super resolution subspace based algorithms, such as ESPRIT\cite{32276}, require the Vandermond property of $\mathbf{A}_l$. 
To address such issues, a virtual receive signal model can be obtained to exploit the potential of sparse arrays, which is based on the second-order statistics of the received signal\cite{5456168}. 
To this end, the covariance matrix of the received signal is given by
\begin{equation} \label{gai}
	\begin{aligned}
		\mathbf{R}  &=\mathbb{E}\left\{\mathbf{y}_l\left[n\right]\mathbf{y}_l^H\left[n\right]\right\} =\mathbf{A}_{l} \mathbf{R}_{l} \mathbf{A}_{l}^H
		+\sigma_n^2 \mathbf{I}_M, \\
	\end{aligned}
\end{equation}
where 
$\mathbf{R}_{l}=\text{diag}\{ |\beta_{K_c+1}|^2P_{K_c+1},\ldots,|\beta_{K}|^2P_{K}\} $ is the covariance matrix of $\mathbf{s}_l\left[n\right]$. Note that in practice, $\mathbf{R}$ is usually computed based on the sample average, i.e.,  $\mathbf{R}=\frac{1}{T}\sum_{n=1}^{T}\mathbf{y}_l\left[n\right]\mathbf{y}_l^H\left[n\right]$, where $T$ in the number of snapshots.  
Then, we vectorize $\mathbf{R}$ to obtain
\begin{equation}
	\label{zRxx}
	\begin{aligned}
		\mathbf{z} & =\operatorname{vec}\left(\mathbf{R}\right)
		=\left(\mathbf{A}_l^* \odot \mathbf{A}_l\right) \mathbf{r}+\sigma_n^2 \overrightarrow{\mathbf{1}},
	\end{aligned}
\end{equation}
where  
$\mathbf{r}=\left[|\beta_{K_c+1}|^2P_{K_c+1},\ldots,|\beta_{K}|^2P_{K}\right]^T$  
can be regarded as the equivalent source signal, and $\sigma_n^2\overrightarrow{\mathbf{1}}=\sigma_n^2
\begin{bmatrix}
	\mathbf{e}_1^T, \mathbf{e}_2^T,\ldots ,\mathbf{e}_M^T
\end{bmatrix}^T$ is the equivalent noise, where $\mathbf{e}_i$ is a zeros vector except 1 at the $i$th position. 
Meanwhile, the conjugate and KR product $\odot$ operations contained in $\left(\mathbf{A}_l^* \odot \mathbf{A}_l\right) $ behave like the difference of the positions of the physical antennas, which forms a virtual array whose antennas locate at $\mathcal{V}=\{\hat{d}_m = \bar{d}_i-\bar{d}_j,1\leq i,j \leq M \}$, where there are $ \left|\mathcal{V}\right| = M_{\mathrm{vir}}$ unique virtual positions. Note that $M_{\mathrm{vir}}\leq M^2$ because there are some difference pairs correspond to the same virtual antenna position and the weight coefficient $ w_m, 1\leq m \leq M_{\mathrm{vir}}$, denotes the number of times $\hat{d}_m$ occurs \cite{5456168}.
After performing virtual array antennas selection and sort, an equivalent steering matrix whose receive antennas locate at the distinct values of $\mathcal{V}$ can be obtained.
For example, it has been proven that sparse MIMO based on nested array with $M$ physical antennas can form a virtual compact ULA with $\frac{M^{2}+2M-2}{2}$ elements when $M$ is even \cite{5456168}. Therefore, the DoFs and array aperture are significantly augmented, rendering it possible to improve the sensing and localization performances.
Note that after creating virtual array, (\ref{zRxx}) becomes a single snapshot signal, thereby spatial smoothing needs to be adopted to recover the rank of the covariance matrix of $\mathbf{z}$ \cite{5456168}. Based on this, estimation algorithms can be performed to obtain the AoAs of the L-UEs in the virtual array domain.

\section{Can Virtual Array Technology Also Benefit Communication?}
Given the remarkable performance of virtual array technology in localization and sensing, we ask the following fundamental question: can virtual array technology also benefit communication for ILAC systems?
To answer this question, in the following, we first present the communication signal processing based on the virtual array, and then demonstrate the impact of virtual array on communication from three aspects, i.e., the resulting information-bearing symbols, SNR comparison, and interference analysis.

\subsection{Communication Signal Processing Based on Virtual Array}
The received signal (\ref{zhangguangjun}) can be equivalently expressed as
\begin{equation}
	\mathbf{y}\left[n\right] = \mathbf{A
	}\mathbf{s}\left[n\right]+\mathbf{n}\left[n\right] ,
\end{equation}
where $\mathbf{A}=[\mathbf{A}_c, \mathbf{A}_l]$ is the steering matrix and $\mathbf{s}\left[n\right]=\begin{bmatrix}
	\mathbf{s}_c^T\left[n\right], \mathbf{s}_l^T\left[n\right]
\end{bmatrix}^T$ denotes the signal vector of UEs.
Besides, $\mathbf{A}_c = \begin{bmatrix}\mathbf{h}_1, \ldots, \mathbf{h}_{K_c}\end{bmatrix}\in\mathbb{C}^{M\times K_c}$ and 
$\mathbf{s}_c\left[n\right]=\left[\sqrt{P_1} x_1\left[n\right], \ldots, \sqrt{ P_{K_c}}x_{K_c}\left[n\right]\right]^T\in\mathbb{C}^{K_c\times 1}$ are signals from C-UEs. 
Similar to the signal processing of virtual array technology in Section \ref{wuyifan}, we compute the conjugate correlation of $\mathbf{y}\left[n\right]$ to get
\begin{equation}\label{duhaoyu}
	\begin{aligned}
		\mathbf{Z}\left[n\right] & = \mathbf{y}\left[n\right]\mathbf{y}^H\left[n\right] \\
		&=\mathbf{A}\mathbf{s}\left[n\right] \mathbf{s}^H\left[n\right]\mathbf{A}^H
		+\mathbf{A}\mathbf{s}\left[n\right]\mathbf{n}^H\left[n\right]\\
		& \quad +\mathbf{n}\left[n\right]\mathbf{s}^H\left[n\right]\mathbf{A}^H
		+\mathbf{n}\left[n\right]\mathbf{n}^H\left[n\right].\\
	\end{aligned}
\end{equation}

For the LoS dominating case, $\mathbf{A}_c$ can be simplified as $\bar{\mathbf{A}}_c=\left[ \beta_{1}\mathbf{a}_1,\ldots,\beta_{K_c}\mathbf{a}_{K_c}\right]\in\mathbb{C}^{M\times K} $, where $\mathbf{a}\left(\theta_i\right)\triangleq \mathbf{a}_i$ for readability and $\beta_k$ represents the LoS path's channel gain of C-UE $k$. Besides, the time index $n$ is omitted for simplicity, and (\ref{duhaoyu}) is converted to
\begin{equation}\label{shenjiajun}
	\begin{aligned}
			\mathbf{Z}  = & \mathbf{a}_k\mathbf{a}_k^{H}|\beta_k|^2 P_k|x_k|^2
			+\sum\nolimits_{i=1,i\neq k}^{K} \mathbf{a}_i\mathbf{a}_i^{H} |\beta_i|^2 P_i|x_i|^2\\
			&+\sum\nolimits_{i=1}^{K}\sum\nolimits_{j=1, j\neq i}^{K} \mathbf{a}_i\mathbf{a}_j^{H} \beta_i\beta_j^H\sqrt{P_iP_j} x_i x_j^H \\
			& +\mathbf{A}\mathbf{s}\mathbf{n}^H
			+\mathbf{n}\mathbf{s}^H\mathbf{A}^H
			+\mathbf{n}\mathbf{n}^H.
		\end{aligned}
\end{equation}
Note that unlike localization signal processing in (\ref{gai}), where the covariance matrix is calculated by the conjugate correlation across different array elements, here we only perform the correlation on the received signals without an expectation operation. This difference arises because localization requires accumulating multiple snapshots to ensure the signal covariance matrix is diagonal, whereas in communication, we are concerned with the information of each transmitted symbol.
Next, $\mathbf{Z}$ is vectorized to
\begin{equation}\label{daizhuoyin}
	\resizebox{0.89\hsize}{!}{$\begin{aligned}
			\mathbf{z}_e & = \mathrm{vec}\left(\mathbf{Z}\right) \\
			& = \left(\mathbf{a}_k^* \otimes \mathbf{a}_k\right) |\beta_k|^2 P_k|x_k|^2 + \sum_{i=1,i\neq k}^{K} \left(\mathbf{a}_i^* \otimes \mathbf{a}_i\right) |\beta_i|^2 P_i |x_i|^2\\
			& \quad +\sum\nolimits_{i=1}^{K}\sum\nolimits_{j=1, j\neq i}^{K} \left(\mathbf{a}_j^* \otimes \mathbf{a}_i\right) \beta_i\beta_j^H \sqrt{P_iP_j}x_i x_j^H
			+ \mathbf{n}_e,
		\end{aligned}$}
\end{equation}
where $\mathbf{n}_e = \mathbf{n}^* \otimes\mathbf{As} + \mathbf{A}^*\mathbf{s}^* \otimes\mathbf{n} + \mathbf{n}^* \otimes\mathbf{n}$.

Note that (\ref{daizhuoyin}) can be regarded as a virtual array signal model, where $\mathbf{z}_e\in\mathbb{C}^{M^2\times 1}$ is the receive signal, $|\beta_k|^2\left(\mathbf{a}_k^*\otimes\mathbf{a}_k\right)$ and $\beta_i\beta_j^H\left(\mathbf{a}_j^*\otimes\mathbf{a}_i\right)$ are the channels, $P_k|x_k|^2$ is the transmit signals, and $\mathbf{n}_e\in\mathbb{C}^{M^2\times 1}$ is the receive noise.
For the steering vector $\mathbf{a}_i^*\otimes\mathbf{a}_i$, the conjugate and Kronecker product $\otimes$ operations behave like the difference of the positions of the physical antennas, which forms a virtual array whose antennas locate at $\mathcal{V}$.

To detect the transmit signal of UE $k$ in the virtual array domain, maximum ratio combining (MRC) beamforming $\mathbf{w}_k = \frac{\mathbf{a}_k^{*}\otimes\mathbf{a}_k}{\|\mathbf{a}_k^{*}\otimes \mathbf{a}_k\|}$ is exploited.
Then the resulting signal for UE $k$ is
\begin{equation} \label{fangchao}
	\resizebox{0.89\hsize}{!}{$\begin{aligned}
		& \eta_k = \mathbf{w}_k^H\mathbf{z} = \left\|\mathbf{a}_k^* \otimes \mathbf{a}_k\right\| |\beta_k|^2|x_k|^2\\ 
		& \quad  + \underbrace{\frac{\left(\mathbf{a}_k^{*}\otimes \mathbf{a}_k\right)^H }{\|\mathbf{a}_k^{*}\otimes \mathbf{a}_k\|}\sum\nolimits_{i=1,i\neq k}^{K}\left(\mathbf{a}_i^{*}\otimes \mathbf{a}_i\right)|\beta_i|^2|x_i|^2 }_{\mathrm{Type\ I\ IUI}}\\
		& \quad + \underbrace{\frac{\left(\mathbf{a}_k^{*}\otimes \mathbf{a}_k\right)^H }{\|\mathbf{a}_k^{*}\otimes \mathbf{a}_k\|}\sum\nolimits_{i=1}^{K}\sum\nolimits_{j=1, j\neq i}^{K} \left(\mathbf{a}_j^* \otimes \mathbf{a}_i\right) \beta_i\beta_j^H x_i x_j^H }_{\mathrm{Type\ II\ IUI}}\\
		& \quad + \frac{\left(\mathbf{a}_k^{*}\otimes \mathbf{a}_k\right)^H }{\|\mathbf{a}_k^{*}\otimes \mathbf{a}_k\|}\left(\mathbf{n}^* \otimes\mathbf{As} + \mathbf{A}^*\mathbf{s}^* \otimes\mathbf{n} + \mathbf{n}^* \otimes\mathbf{n}\right).
	\end{aligned}$}
\end{equation}
We can infer from (\ref{fangchao}) that there are several forms of interference and noise, where the interference is divided into two categories, namely Type I and Type II IUI, respectively. 
Moreover, we find that the information-bearing signal $x_k$ only appears in forms of $|x_k|^2$, causing the communication symbols also undergo selfcorrelation. It makes the information-bearing signal $x_k$ lose the phase information, which is detrimental to communication scenarios.

\subsection{SNR comparison}
To assess the impact of virtual array technology on SNR, we consider a simplified single LoS UE case.
The received signal is
\begin{equation}
	\mathbf{y}\left[n\right]=\mathbf{a}\beta\sqrt{P}x\left[n\right]+\mathbf{n}\left[n\right],
\end{equation}
where $\mathbf{a}$ is the steering vector of sparse array.
$x\left[n\right]$ and $\beta$ are the transmit symbol and pathloss, respectively.
It is well known that the maximum SNR after the optimal MRC beamforming is $\mathrm{SNR}_{phy} = \frac{MP|\beta|^2}{\sigma^2}$, where the array beamforming gain is $M$.
The virtual array based received signal model can be degraded from (\ref{daizhuoyin}) when $K=1$ and $K_c=1$, which is
\begin{equation}\label{zhouting}
	\begin{aligned}
		\mathbf{z}_s\left[n\right] & = \left(\mathbf{a}^*\otimes\mathbf{a}\right)  |\beta|^2|Px\left[n\right]|^2 
		+ \left(\mathbf{n}^*\otimes\mathbf{a}\right) \beta\sqrt{P}x\left[n\right]\\
		& \quad + \left(\mathbf{a}^*\otimes\mathbf{n}\right)\beta^*\sqrt{P}x^*\left[n\right]
		+ \mathbf{n}^*\left[n\right]\otimes\mathbf{n}\left[n\right].
	\end{aligned}	
\end{equation}
Performing MRC beamforming $\mathbf{w}_{s}=\frac{\mathbf{a}^*\otimes\mathbf{a}}{\|\mathbf{a}^*\otimes\mathbf{a}\|}=\frac{1}{M}\mathbf{a}^*\otimes\mathbf{a}$ on the virtual array, the resulting signal is
\begin{equation} \label{pengyuan}
	\begin{aligned}
		&\eta_{s} = \mathbf{w}_{s}^H\mathbf{z}_s\left[n\right] = M P|\beta|^2|x\left[n\right]|^2
		+\left(\mathbf{a}^T \mathbf{n}^*\left[n\right]\right) \beta^*\sqrt{P} x^*\left[n\right] \\
		& \quad \quad \quad \quad  +\left(\mathbf{n}^T\left[n\right] \mathbf{a}^*\right)\beta \sqrt{P} x\left[n\right]  
		+\frac{\mathbf{a}^H \mathbf{n}\left[n\right] \mathbf{n}^H\left[n\right] \mathbf{a}}{M}.
	\end{aligned}
\end{equation}


Then the SNR of the received signal $\left|x_k\right|^2$ is
\begin{equation}\label{baideng}
	\resizebox{0.89\hsize}{!}{$
		\begin{aligned}
			&\mathrm{SNR}_{vir} = \\ &\frac{E\left\{M^2  P^2 |\beta|^4 |x\left[n\right]|^4\right\}}{E\left\{\left|\left(\mathbf{a}^T \mathbf{n}^*\left[n\right]\right)\beta^* \sqrt{P} x^*\left[n\right] + \left(\mathbf{n}^T\left[n\right] \mathbf{a}^*\right)\beta \sqrt{P} x\left[n\right] +\frac{\mathbf{a}^H \mathbf{n}\left[n\right] \mathbf{n}^H\left[n\right] \mathbf{a}}{M}\right|^2\right\}}, \\
		\end{aligned}$}
\end{equation}
which is proved to satisfy
\begin{equation}\label{mahuateng}
	\mathrm{SNR}_{vir} \leq \mathrm{SNR}_{phy}.
\end{equation}
\begin{IEEEproof}
	See Appendix \ref{wuzhenyang}.
\end{IEEEproof}
This implies that even though a virtual array with larger aperture is formed, the additional interference and noise in (\ref{pengyuan}) leads to degradation in SNR.

\subsection{Virtual Array Beam Pattern Analysis}
Based on the so-called Type I and II IUI in (\ref{fangchao}), we analyze the sparse MIMO's interference suppressing capability through the following beam patterns. 
\subsubsection{Type I IUI Beam Pattern} $ G_{\mathrm{vir}}\left(\Delta_{ki};\mathcal{V}\right) \triangleq \frac{ \left|\left(\mathbf{a}_k^{*}\otimes \mathbf{a}_k\right)^H\left(\mathbf{a}_i^{*}\otimes \mathbf{a}_i\right)\right|^2 }{\|\mathbf{a}_k^{*}\otimes \mathbf{a}_k\|^2 \|\mathbf{a}_i^{*}\otimes \mathbf{a}_i\|^2} $ reflects the interference of the signal $|x_i|^2|\beta_i|^2$ from UE $i$ with $i\neq k$. 
\begin{equation} \label{lvxiaoxiao}
	\begin{aligned}
		& G_{\mathrm{vir}}\left(\Delta_{ki}, \mathcal{V}\right)  = \frac{1}{M^4}\left|\sum_{m=1}^{M_{\mathrm{vir}}}w_m e^{j\pi\hat{d}_m\Delta_{ki}}\right|^2.
	\end{aligned}
\end{equation}

\subsubsection{Type II IUI Beam Pattern} $ H_{\mathrm{vir}}\left(\Delta_{kj},\Delta_{ki}; \mathcal{V}\right) \triangleq  \frac{\left|\left(\mathbf{a}_k^{*}\otimes \mathbf{a}_k\right)^H\left(\mathbf{a}_j^{*}\otimes \mathbf{a}_i\right)\right|^2}{\|\mathbf{a}_k^{*}\otimes \mathbf{a}_k\|^2 \|\mathbf{a}_j^{*}\otimes \mathbf{a}_i\|^2}$ reflects the interference of the signal $x_ix_j^H\beta_i\beta_j^H$ from UE $i$ and $j$ with $i\neq j$, which can also be rewritten as 
\begin{equation} \label{huangjijie}
	\resizebox{0.89\hsize}{!}{$
		\begin{aligned}
			&H_{\mathrm{vir}}\left(\Delta_{ki},\Delta_{kj}; \mathcal{V}\right) 
			= \frac{1}{M^4}\left|\left(\mathbf{a}_k^{T}\otimes \mathbf{a}_k^H\right)	\left(\mathbf{a}_i^{*}\otimes \mathbf{a}_j\right)\right|^2\\
			&= \frac{1}{M^4}\left|\left(\mathbf{a}_k^{T}\mathbf{a}_i^{*}\right)\otimes\left(\mathbf{a}_k^H\mathbf{a}_j\right)\right|^2
			= G\left(\Delta_{ki};\mathcal{D}\right)G\left(-\Delta_{kj};\mathcal{D}\right).
		\end{aligned}$}
\end{equation}

%
As shown in Fig. \ref{fig5_BP_IUI_1} and Fig. \ref{zhaochunming}(a), we take the nested array with $\left(N_1,N_2\right)=\left(4,4\right)$ as an example and plot the Type I and II IUI beam patterns. It can be observed that, compared to compact ULA, sparse array forms narrower main lobe beam width in both the Type I IUI and Type II IUI beam patterns. However, it only improves the interference suppressing capability of Type I IUI, while the capability of suppressing Type II IUI is compromised. There are two reasons for this phenomenon: Firstly, in Type II IUI, deeper null steering points locate only at the four corners near the main lobe, approximately $-33.16$ dB, while the side lobes along the coordinate axes are relatively high (shown in the red dashed box), around $-10.93$ dB. Secondly, compared to the $K-1$ terms of Type I IUI, the number of Type II IUI terms is $K\left(K-1\right)$, and as the number of users increases, it becomes more difficult to suppress Type II IUI, resulting in a degradation of communication performance.

However, the aforementioned virtual array technology does not eliminate the impact of weight coefficients in (\ref{lvxiaoxiao}). This results in the appearance of some dominant grating lobes in the beam pattern as shown in Fig. \ref{fig5_BP_IUI_1}, which affects the Type I IUI interference suppressing capability.
\subsubsection{Type I IUI Beam Pattern of Selected Virtual Array} 
If a vitual array antennas selection is performed on $\mathcal{V}$ to eliminate these weight coefficients, a virtual compact ULA $\mathcal{V}_s$ with $\left|\mathcal{V}_s\right| = M_{\mathrm{svir}}$ can be formed, where $M_{\mathrm{svir}}$ is the number of antennas of the continuous part of $\mathcal{V}$, thereby $\mathcal{V}_s \subseteq \mathcal{V}$. Note that the nested array is able to form a virtual compact ULA with $\mathcal{V}_s=\mathcal{V}$ \cite{5456168}, but for co-prime array, there may exist holes in the virtual array, leading to $\mathcal{V}_s \subset \mathcal{V}$ \cite{vaidyanathanSparseSensingCoPrime2011b}.
After that, the corresponding MRC beamformer of UE $k$ is $\check{\mathbf{w}}_k = \frac{\check{\mathbf{a}}_k}{\left\|\check{\mathbf{a}}_k\right\|}$, where $\check{\mathbf{a}}_k = \begin{bmatrix}	1,e^{j\pi\check{d}_2\sin\theta_k},...,e^{j\pi\check{d}_{M_{\mathrm{svir}}}\sin\theta_k} \end{bmatrix}$ is the steering vector.
Therefore, the Type I IUI beam pattern of the selected virtual array $\mathcal{V}_s$ is
\begin{equation} \label{xutianheng}
	\resizebox{0.89\hsize}{!}{$\begin{aligned}
		& G_{\mathrm{svir}}\left(\Delta_{ki};\mathcal{V}_s\right)  \\
		& =  \frac{1}{M_{\mathrm{svir}}^2}\left|\sum\nolimits_{m=1}^{M_{\mathrm{svir}}} e^{j\pi\check{d}_m\Delta_{ki}}\right|^2
		= \frac{1}{M_{\mathrm{svir}}^4}\left|\frac{\sin (\frac{\pi}{2} M_{\mathrm{svir}}\Delta)}{\sin (\frac{\pi}{2}\Delta)}\right|^2.
	\end{aligned} $}
\end{equation}

\subsubsection{Type II IUI Beam Pattern of Selected Virtual Array} Correspondingly, the Type II IUI beam pattern of the selected virtual array becomes
\begin{equation}\label{xuhongbiao}
	\resizebox{0.89\hsize}{!}{$\begin{aligned}
		& H_{\mathrm{svir}}\left(\Delta_{ki},\Delta_{kj}; \mathcal{V}_s\right)  = \frac{1}{M_{\mathrm{svir}}^2}
		\left|\sum\nolimits_{m=1}^{M_{\mathrm{svir}}} e^{j\pi\left(p_m\Delta_{ki}-q_m\Delta_{kj}\right)}\right|^2, \\
	\end{aligned}$}
\end{equation}
where $p_m,q_m \in \{\bar{d}_m\}$ and $p_m-q_m = \hat{d}_m$. 

\begin{figure}[h]
	\centering
	\includegraphics[width = 6cm]{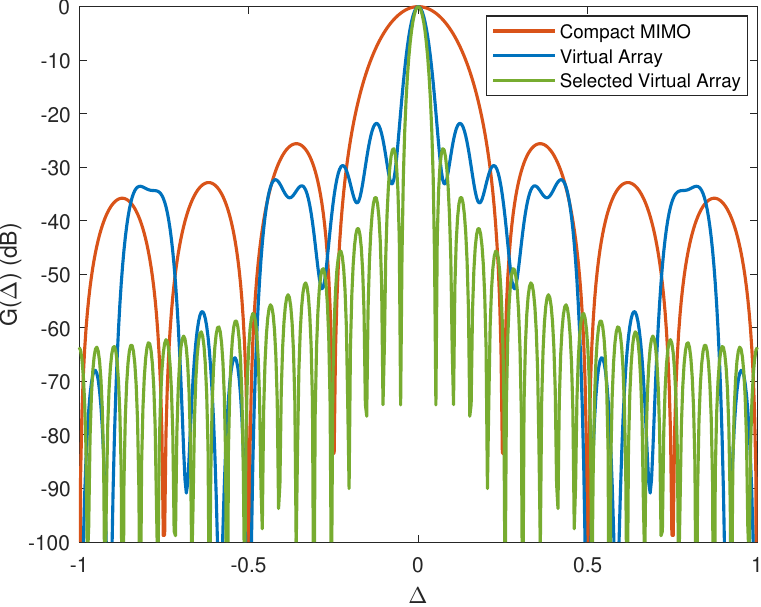}
	\caption{The Type I IUI beam pattern of compact MIMO, sparse MIMO based on nested array's virtual array and selected virtual array, i.e., $G\left(\Delta;\mathcal{D}\right)$,
		$G_{\mathrm{vir}}\left(\Delta;\mathcal{V}\right)$, $G_{\mathrm{vir}}\left(\Delta;\mathcal{V}_s\right)$, respectively.}	
	\label{fig5_BP_IUI_1}	
\end{figure}

\begin{figure}[h]
	\centering
	\begin{subfigure}{0.49\linewidth}
		\centering
		\includegraphics[width=0.9\linewidth]{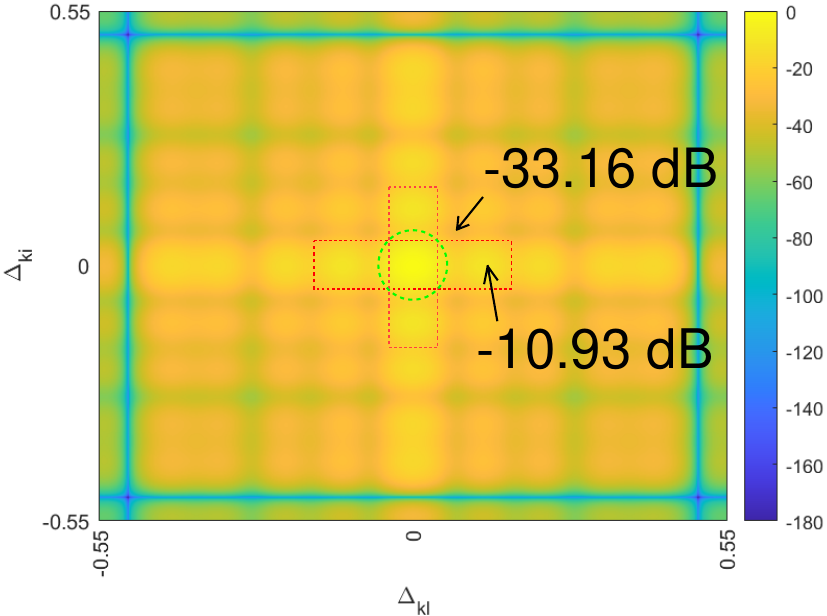}
		\caption{$H_{\mathrm{vir}}\left(\Delta_{kl},\Delta_{ki};\mathcal{V}\right)$.}
		\label{crossIUI}
	\end{subfigure}
	\centering
	\begin{subfigure}{0.49\linewidth}
		\centering
		\includegraphics[width=0.9\linewidth]{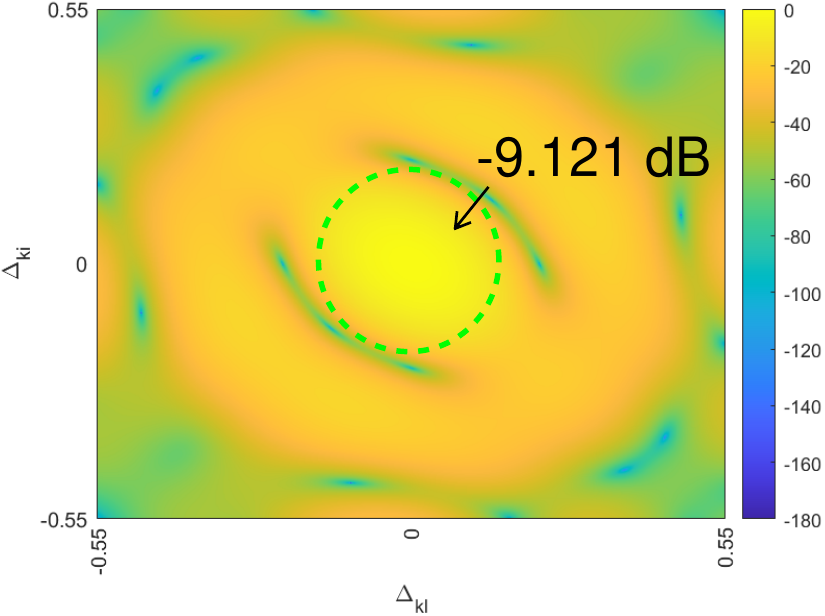}
		\caption{$H_{\mathrm{svir}}\left(\Delta_{kl},\Delta_{ki};\mathcal{V}_s\right)$.}
		\label{crossIUI_sel}
	\end{subfigure}
	\caption{The Type II IUI beam pattern of virtual array of nested based sparse MIMO.}
	\label{zhaochunming}
\end{figure}

As shown in Fig. \ref{fig5_BP_IUI_1}, after performing antenna selection for the virtual array, the main lobe of the Type I IUI beam pattern becomes narrower, and there are no irregular grating lobes, resulting in better Type I IUI suppressing. However, the main lobe region of the Type II IUI beam pattern becomes much larger, which further aggregate the multi-user interference, as shown in Fig. \ref{zhaochunming}(b). 

In summary, the virtual array technology for communication will not only destroy the phase information of the transmitted symbols, degrade the SNR of received signal, but also aggravate the interference. These are the reasons why it is difficult to be applied in communication.
Therefore, in this paper, we propose the hybrid processing for sparse MIMO based ILAC, i.e., physical array based communication while virtual array based localization. To this end, we need to first characterize the beam pattern of the sparse arrays in the physical array domain.

\section{Hybrid Processing for Sparse MIMO Based ILAC} \label{liyue}
In this section, physical array based communication signal processing is firstly presented. Then the beam pattern of the sparse MIMO based on nested and co-prime arrays are analyzed. The extracted beam pattern information, including $\mathrm{BW}$, $\mathrm{PLMR}$ and $\mathrm{SLH}$, demonstrate the potential of sparse MIMO in providing better spatial resolution compared to its compact counterpart with the same number of physical array elements.

\subsection{Physical Array Based Communication Signal Processing}
Previous works \cite{10404673, 10465094} have shown that, even in the presence of grating lobes in the beam pattern of its physical array, sparse MIMO still achieve better communication performance compared to its conventional counterpart, especially in scenarios where UEs are densely distributed. 

To detect the signal of C-UE $k$, a linear receive beamforming vector $\mathbf{v}_k \in \mathbb{C}^{M \times 1}$ with $\left\|\mathbf{v}_k\right\|=1$ is used and the resulting signal is
\begin{equation}
	\resizebox{0.89\hsize}{!}{$\begin{aligned}
			y_k=\mathbf{v}_k^H \mathbf{y}\left[n\right]=\mathbf{v}_k^H \mathbf{h}_k \sqrt{P_k} x_k\left[n\right]+\mathbf{v}_k^H \sum_{i=1, i \neq k}^{K} \mathbf{h}_i \sqrt{P_i} x_i\left[n\right]
			+\mathbf{v}_k^H \mathbf{n}\left[n\right].\\
		\end{aligned}$}
\end{equation}	

Then the signal-to-interference-plus-noise ratio (SINR) of C-UE $k$ is
\begin{equation}\label{masike}
	\begin{aligned}
		\gamma_k=\frac{P_k\left|\mathbf{v}_k^H \mathbf{h}_k\right|^2}{\sum_{i=1, i \neq k}^K P_i\left|\mathbf{v}_k^H \mathbf{h}_i\right|^2
			+\sigma^2}.
	\end{aligned}
\end{equation}
Therefore, the achievable rate of C-UE $k$ can be expressed as 
\begin{equation}
	\label{wangsihan}
	R_k=\log _2(1+\gamma_k) .
\end{equation}

From (\ref{masike}), it can be observed that the SINR of C-UE $k$ not only depends on the beamforming gain introduced by $\left|\mathbf{v}_k\mathbf{h}_k\right|^2$ in numerator, but also relates to the interference induced by $\left|\mathbf{v}_k\mathbf{h}_i\right|^2$ in the denominator.
For a given number of antennas $M$ at BS, the optimal beamforming gain with the value $M$ can be realized by the MRC beamformer $\mathbf{v}_k=\frac{\mathbf{h}_k}{\|\mathbf{h}_k\|}$. 
Moreover, given the distribution of the users, the SINR mainly depends on the array structure $\mathcal{D}$ of the sparse MIMO, which is contained within the array steering vector.

For the special LoS-dominating case, we have $L_k=1$. In this case, the channel vector reduces to
$\mathbf{h}_k=\beta_{k} \mathbf{a}\left(\theta_{k}\right)$, where $\beta_k$ and $\theta_k $ are the channel gain and AoA of UE $k$, respectively.
Thereby, the corresponding SINR can be simplified as
\begin{equation}\label{bf_methods}
	\resizebox{0.89\hsize}{!}{$\begin{aligned}
			\gamma_{k} 
			= \frac{P_k\left|\beta_k\right|^2M}
			{
				\sum\limits_{i=1, i \neq k}^{K} P_i M \left|\beta_i\right|^2 \left|\frac{\mathbf{a}_k^H \mathbf{a}_i}{\left\|\mathbf{a}_k\right\|\left\|\mathbf{a}_i\right\|}\right|^2 
				+\sigma^2
			}
			= \frac{\bar{P}_k M}
			{M \sum\limits_{i=1, i \neq k}^{K} \bar{P}_i G\left(\Delta_{ki};\mathcal{D}\right)	+1},
		\end{aligned}$}
\end{equation}
where $\bar{P}_k \triangleq \frac{\left|\beta_k\right|^2 P_k}{\sigma^2}$ is the receive $\mathrm{SNR}$.
Note that the subscripts of $\Delta_{ki}$ are omitted in the following for brevity. 
In contrast to compact MIMO, which has evenly spaced null steering points and smoothly descending side lobes, the beam pattern of sparse MIMO exhibits significant irregularity corresponding to the sparse array structures due to the grating lobes.
Conventional metrics like first null beam width or half power beam width can no longer accurately assess its communication performance.
Therefore, four typical beam pattern matrices of sparse MIMO are provided here.
\begin{enumerate}
	\item First local minimum point ($\mathrm{FLMP}$), which is denoted by $\Delta_{min}$. The $\mathrm{FLMP}$ of the sparse MIMO's beam pattern $G\left(\Delta;\mathcal{D}\right)$ is defined as the smallest local minimum point on the positive semi-axis, i.e.,
	\begin{equation}
		\Delta_{min}= \min\{\Delta \mid \Delta\in\{\Delta_{i}^{loc}\}_{i=1}^{n}\},
	\end{equation}
	where $\{\Delta_{i}^{loc}\}_{i=1}^{n_{\mathrm{loc}}}$ are the local minimum points of $G(\Delta;\mathcal{D})$.
	
	\item $\mathrm{BW}$. The $\mathrm{BW}$ of sparse MIMO is defined as $\text{BW}=2\Delta_{min}$, which intuitively reflects the spatial resolution of sparse MIMO. 
	
	\item $\mathrm{PLMR}$. $\mathrm{PLMR}$ characterizes the ability to suppress interference, which is be defined by the ratio of $G(0)$ to the beam gain of the $\mathrm{FLMP}$, i.e.,
	\begin{equation}
		\mathrm{PLMR} = \frac{G\left(0\right)}{G\left(\Delta_{min}\right)} = \frac{1}{G\left(\Delta_{min}\right)},
	\end{equation}
	where $G\left(0\right)=(N_1+N_2)^2/M^2=1$.
	
	\item $\mathrm{SLH}$.
	Due to the presence of grating lobes in the beam pattern of sparse arrays, the positions and the beam gain of these lobes are also important metrics reflecting the IUI situation of communication and the possible spurious peaks of localization or sensing. 
\end{enumerate}

\subsection{Nested Array Beam Pattern Analysis}
By substituting $\mathcal{D}_{\mathrm{na}}$ into (\ref{telangpu}), the physical array based beam pattern of the nested array $\left(N_1, N_2\right)$ can be obtained as
\begin{equation}
	\label{qianhua}
	\resizebox{0.88\hsize}{!}{$\begin{aligned}
			& G\left(\Delta; \mathcal{D}_{\mathrm{na}}\right) \\
			& =\frac{1}{M^2}\left|
			\frac{\sin (\frac{\pi}{2} N_1\bar{d}_{in}\Delta)}{\sin (\frac{\pi}{2}\bar{d}_{in}\Delta)}
			+e^{j\frac{\pi}{2}N_2\bar{d}_{ou}\Delta}
			\frac{\sin(\frac{\pi}{2} N_2\bar{d}_{ou}\Delta)}{\sin(\frac{\pi}{2} \bar{d}_{ou}\Delta)}\right|^2\\
		\end{aligned}$},
\end{equation}
where $\bar{d}_{in}=1$ and $\bar{d}_{ou}=N_1+1$ are the adjacent element spacing of inner and outer subarrays on the uniform grid, respectively.
Owning to the non-uniform architecture of physical array, the corresponding beam pattern of nested array $G\left(\Delta; \mathcal{D}_{\mathrm{na}}\right)\triangleq G_{\mathrm{na}}$ generates grating lobes and dominating side lobes.
To gain some insights, (\ref{qianhua}) is simplified as
\begin{equation}\label{yanhongping}
	\begin{aligned}
		G_{\mathrm{na}}&=\frac{1}{M^2}\left|\lambda\right|^2
		=\frac{1}{M^2}\left|f+e^{j\Phi}g\right|^2\\
		&=\frac{1}{M^2}\left(f^2+g^2+2fg\cos(\Phi)\right),\\
	\end{aligned}
\end{equation}
where $\lambda=f+e^{j\Phi}g$, $f=\frac{\sin (\frac{\pi}{2} N_1\bar{d}_{in}\Delta)}{\sin (\frac{\pi}{2}\bar{d}_{in}\Delta)}$, $g=\frac{\sin(\frac{\pi}{2} N_2\bar{d}_{ou}\Delta)}{\sin(\frac{\pi}{2} \bar{d}_{ou}\Delta)}$ and $\Phi=\frac{\pi}{2}N_2\bar{d}_{ou}\Delta$. It is found that the main lobe of $G_{\mathrm{na}}$ can be approximately regarded as the combination of the main lobes of $f$ and $g$ except a complex coefficient $e^{j\Phi}$. 
Due to the highly non-linearity of $G_{\mathrm{na}}$, we analyze the contributions of each sub-term to the beam pattern in (\ref{yanhongping}). Firstly, we extract the first null point ($\mathrm{FNP}$) information of each component.

\subsubsection{$\mathrm{FNP}$}
For $f$, let $\frac{\pi}{2} N_1\bar{d}_{in}\Delta=\pi$, yielding
$\Delta_1=\frac{2}{N_1\bar{d}_{in}}=\frac{2}{N_1}$;
For $g$, let $\frac{\pi}{2} N_2 \bar{d}_{ou}\Delta=\pi$, yielding
$\Delta_2=\frac{2}{N_2\bar{d}_{ou}}=\frac{2}{(N_1+1)N_2}$;
For $\cos(\Phi)$, let $\frac{\pi}{2} N_2 \bar{d}_{ou}\Delta=\frac{\pi}{2}$, yielding
$\Delta_3=\frac{1}{N_2\bar{d}_{ou}}=\frac{1}{(N_1+1)N_2}$, and these $\mathrm{FNP}$s satisfy $\Delta_1=\frac{\left(N_1+1\right)N_2}{N_1}\Delta_2$, $\Delta_2 =2\Delta_3 $ and $\Delta_3  < \Delta_2  < \Delta_1$.

\subsubsection{BW}
The lower and upper bounds of $\mathrm{FLMP}$ can be obtained, which are related to the aforementioned $\mathrm{FNP}$ information. 
\begin{theorem}
	\label{wuyulin}
	The FLMP $\Delta_{min}$ of nested array is bounded by
	\begin{itemize}
		\item When $ 2\leq N_2\leq N_{th}$,
		\begin{equation} \label{xuxuan}
			\frac{2(N_2-1)}{(N_1+1)N_2} \leq\Delta_{min} \leq \frac{2}{N_1+1}.
		\end{equation}
		\item When $ N_{th}< N_2\leq N_{ap}$,
		\begin{equation}
			\frac{1}{(N_1+1)N_2} \leq\Delta_{min} \leq \frac{2}{(N_1+1)N_2}. 
		\end{equation}
		\item When $ N_2> N_{ap}$,
		\begin{equation} \label{pengpeiao}
			\Delta_{int}\leq\Delta_{min} \leq\frac{2}{(N_1+1)N_2},
		\end{equation}
	\end{itemize}
	where $\Delta_{int}$ is a unique solution to the equation $\cos(\Phi)=\frac{g(\Phi)}{2f(0)}$ for $\Phi\in[\frac{\pi}{2},\pi]$. Besides, $N_{th}$ and $N_{ap}$ are two threshold values of $N_2$ for a given $N_1$, where
	\begin{equation*}
		\resizebox{0.85\hsize}{!}{$
			N_{th}= \left\{
			\begin{aligned}
				&1,& N_1<7, \\
				&\max\{N_2 \mid G^{\prime}\left(\Phi\right) < 0,  0 \leq \Phi \leq \pi  \}, &N_1\geq 7,\\
			\end{aligned}
			\right.$}
	\end{equation*}
	and $N_{ap} \approx \lfloor \sqrt{\frac{10N_1^2}{N_1+1}} \rfloor$.
\end{theorem}

Note that when the outer subarray is much smaller than the inner subarray, i.e., $N_2\leq N_{th}\ll N_1$, the BW of nested array mainly depends on the inner subarray in the asymptotic case, i.e., $\mathrm{BW}\rightarrow  \mathrm{BW}_{in}=2\Delta_1$ for $N_1\rightarrow+\infty$.
By contrast, when the outer subarray is large enough, i.e., $ N_2\gg N_{ap}$, the $\mathrm{BW}$ is determined by the outer subarray $\mathrm{BW}\rightarrow  \mathrm{BW}_{ou}=2\Delta_2$ when $N_2\rightarrow+\infty$. 

\subsubsection{PLMR}
It is worth noting that the $\mathrm{FLMP}$ $\Delta_{min}$ does not have a closed-form expression. Thus, lower bound of PLMR is used to approximate the actual value, i.e.,
\begin{lemma}
	\label{fengyingqi}
	$\mathrm{PLMR}$ of a nested array $\left(N_1, N_2\right)$ is  
	\begin{itemize}
		\item When $ 2\leq N_2\leq N_{th}$, 
		$
		\mathrm{PLMR }\geq \max\{\frac{1}{P_4}, \frac{1}{P_5}\}.
		$
		\item When $ N_{th}\leq N_2\leq N_{ap}$, 
		$
		\mathrm{PLMR }\geq \max\{ \frac{1}{P_{3}}, \frac{1}{P_{2}}\}.
		$
		\item When $ N_2\geq N_{ap}$,
		$
		\mathrm{PLMR }\geq \max\{ \frac{1}{P_{int}}, \frac{1}{P_{2}}\},
		$
	\end{itemize}
	where $P_{1}= G(\Delta_1 )=\frac{1}{M^2}g^2(\Delta_1 )$, 
	$P_{2} = G(\Delta_2 )=\frac{1}{M^2}f^2(\Delta_2 )$, 
	$P_{3} = G(\Delta_3 )=\frac{1}{M^2}\left(f^2(\Delta_3 )+g^2(\Delta_3 )\right)$, 
	$P_{4}=G\left(\left(N_2-1\right)\Delta_2\right)=\frac{1}{M^2}f^2\left(\left(N_2-1\right)\Delta_2\right)$, 
	$P_{5} =G\left(N_2\Delta_2 \right)=\frac{1}{M^2}\left(f^2\left(N_2\Delta_2 \right)+N_2^2-2N_2f\left(N_2\Delta_{2}\right)\right)$,
	$P_{int}= G(\Delta_{int})$.
\end{lemma}

\subsubsection{SLH} Here, the characteristics of grating lobes and side lobes based on nested arrays are approximately provided.
\begin{theorem}
	\label{qianxiufeng}
	Given $N_2> N_{ap}$, the dominating side lobes appear at 
	$
	\Delta_{\mathrm{s, n}} \approx \frac{2n}{N_1+1}, n = \pm1, \pm2,\ldots, \pm  N_1, 
	$
	and these lobes have the similar height, which is
	$
	\mathrm{SLH } \approx \frac{\left(N_2-1\right)^2}{M^2}.
	$
\end{theorem}

Otherwise, when $N_2 \leq N_{ap}$, the beam pattern of nested array depends mainly on the inner subarray because the amplitude of the outer subarray is too small compared with the inner one, submerging the grating lobes of the outer subarray in the beam pattern of nested array. Therefore, the $\mathrm{SLH}$ of this case is similar to the inner subarray and the details are omitted here. 
\begin{figure}[htbp]
	\centering
	\includegraphics[width = 6cm]{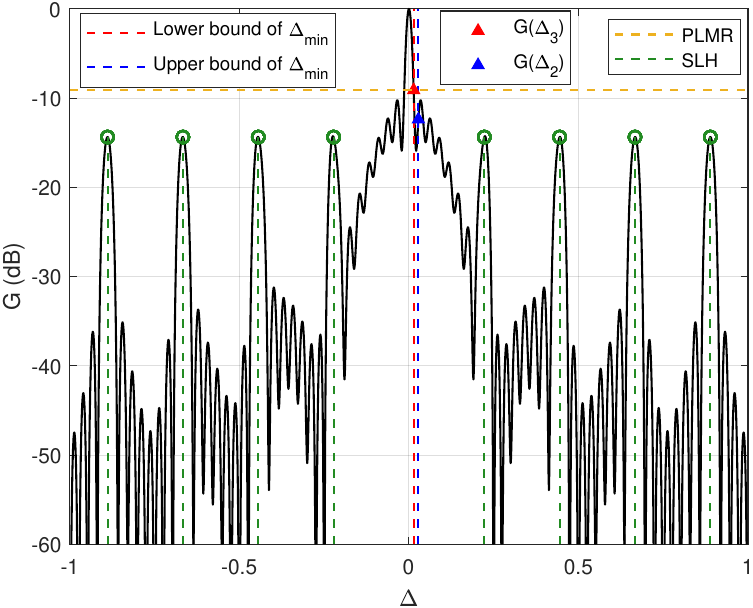}
	\caption{Beam pattern of nested array with $\left(N_1,N_2\right)=\left(8,8\right)$.}
	\label{beampattern_na}
\end{figure}

Fig. \ref{beampattern_na} validates the effectiveness of the proposed nested array beam pattern characteristics, which contain $\mathrm{BW}$, $\mathrm{PLMR}$ and $\mathrm{SLH}$ metrics.
Based on Theorem \ref{wuyulin}, we can get $N_{ap}=8$ for $N_1=8$. 
Firstly, it is observed that the beam pattern can be approximated as the linear combination of the beam patterns of two subarrays when $N_2\geq N_{ap}$. 
Secondly, the main lobe of the inner subarray forms lots of vibration due to the beam pattern of the outer subarray. Especially, the beam pattern of nested array forms main lobe within this region, and the beam width is $\mathrm{BW}\leq\frac{4}{(N_1+1)N_2} $ based on (\ref{pengpeiao}). The $\mathrm{BW}$ is much narrower than that of its compact counterpart, whose main lobe beam width is $\mathrm{BW}_c=\frac{4}{M}$, because $M=N_1+N_2\ll (N_1+1)N_2$. This enables sparse MIMO to achieve higher spatial resolution and ability to suppress IUI. 
Lastly, the side lobes can be captured by Theorem \ref{qianxiufeng}. Thereby, side lobes suppression based on the beamforming optimization and even user grouping can be realized to mitigate IUI introduced by these side lobes \cite{xinruiLiSparseMIMO}.

\subsection{Co-prime Array Beam Pattern Analysis}
By substituting $\mathcal{D}_{\mathrm{cp}}$ into (\ref{telangpu}), the physical array based beam pattern of the co-prime array $\left(M_1, M_2\right)$ can be obtained as 
\begin{equation} \label{xuxiaoli}
	\begin{aligned}
		G\left(\Delta; \mathcal{D}_{\mathrm{cp}}\right)
		&=\frac{1}{M^2}\left|
		e^{j\pi\frac{1}{2}(M_1-1)M_2\Delta}
		\frac{\sin(\pi M_1M_2\Delta)}{\sin(\frac{1}{2}\pi M_2\Delta)}\right. \\
		& \quad \left. + \frac{\sin(\frac{1}{2}\pi M_1(M_2-1)\Delta)}{\sin(\frac{1}{2}\pi M_1\Delta)}
		\right|^2. \\
	\end{aligned}
\end{equation}
To get some insights, (\ref{xuxiaoli}) is simplified to
\begin{align}
	G_{\mathrm{cp}}	& =\frac{1}{M^2}\left|	e^{j\Phi}I_1 + I_2	\right|^2 \label{yangyuhang} \\ 
					& =\frac{1}{M^2}\left(	I_1^2 + I_2^2 + 2I_1I_2\cos \Phi\right), \label{liuyuchen}
\end{align}
where $I_1=\frac{\sin(\pi M_1M_2\Delta)}{\sin(\frac{1}{2}\pi M_2\Delta)}$, $I_2=\frac{\sin(\frac{1}{2}\pi M_1(M_2-1)\Delta)}{\sin(\frac{1}{2}\pi M_1\Delta)}$ and $\Phi=\frac{1}{2}\pi(M_1-1)M_2\Delta$.
The main lobe of $G_{\mathrm{cp}}$ can be approximately regarded as the combination of the main lobes of $I_1$ and $I_2$ except a complex coefficient $e^{j\Phi}$. 

\subsubsection{FNP}
For $I_1$, let $\pi M_1M_2\Delta_1=\pi$, yielding $\Delta_1=\frac{1}{M_1M_2}$; 
For $I_2$, let $\frac{1}{2}\pi M_1\left(M_2-1\right)\Delta_2=\pi$, yielding $\Delta_2 = \frac{2}{M_1\left(M_2-1\right)}$; 
For $\cos\Phi$, let $\frac{1}{2}\pi(M_1-1)M_2\Delta_3 = \frac{\pi}{2}$, yielding $\Delta_3 = \frac{1}{\left(M_1-1\right)M_2}$. 
Besides, $\Delta_a \approx \frac{3}{2}\Delta_1=\frac{3}{2M_1M_2}$ is a approximation of the $\mathrm{FLMP}$ of $I_1$. 
Therefore, a generalized inequality concerning these points can be obtained, $\Delta_1<\Delta_3< \Delta_a < 2\Delta_1<\Delta_2$, where $\Delta_3< \Delta_a$ holds when $M_1\geq 4$, $\Delta_3\leq 2\Delta_1$ holds when $M_1\geq 2$, and $\Delta_3\rightarrow\Delta_1$ when $M_1M_2\rightarrow \infty$.

\subsubsection{BW}
The $\mathrm{FLMP}$'s bounds of co-prime array can be obtained based on the aforementioned $\mathrm{FNP}$s. 
\begin{theorem}
	\label{liulichu}
	For a given $M_1$, there exists two threshold values $M_{th,1}$ and $M_{th,2}$. The $\mathrm{FLMP}$ $\Delta_{min}$ is bounded by
	\begin{itemize}
		\item When $ M_1 < M_2\leq M_{th,1}$,
		\begin{equation}\label{yexiaoying}
			\frac{1}{M_1M_2} \leq\Delta_{min} \leq \frac{1}{\left(M_1-1\right)M_2}.
		\end{equation}
		
		\item When $ M_{th,1} < M_2\leq M_{th,2}$,
		\begin{equation} \label{wangxi}
			\frac{1}{\left(M_1-1\right)M_2}\leq\Delta_{min} \leq \frac{3}{2M_1M_2}.
		\end{equation}
		
		\item When $ M_{th,2}< M_2$,
		\begin{equation} \label{tangwen}
			\frac{2}{M_1M_2} \leq\Delta_{min} \leq \frac{2}{M_1(M_2-1)}. 
		\end{equation}
	\end{itemize}
	Note that when $M_1\geq 14$, $M_{th,1}$ degenerates to $M_1$.
\end{theorem}
\begin{IEEEproof}
	See appendix \ref{lvjiangbin}
\end{IEEEproof}
\subsubsection{PLMR} Based on the $\mathrm{BW}$ analysis, the $\mathrm{PLMR}$ of co-prime array can be obtained as follows.
\begin{lemma}
	\label{wenchaokai}           
	The $\mathrm{PLMR}$ of co-prime array is bounded by
	\begin{itemize}
		\item When $ M_1< M_2\leq M_{th,1}$,
		$
		\mathrm{PLMR} \geq \max \left\{\frac{1}{Q_1}, \frac{1}{Q_3} \right\}.
		$
		
		\item When $ M_{th,1} < M_2\leq M_{th,2}$,
		$
		\mathrm{PLMR} \geq  \frac{1}{Q_3}. 
		$
		
		\item When $ M_{th,2}< M_2$,
		$
		\mathrm{PLMR} \geq \max \left\{\frac{1}{Q_2}, \frac{1}{Q_4} \right\}, 
		$
	\end{itemize}
\end{lemma}
where $Q_1 = G_{\mathrm{cp}} \left(\Delta_1\right) = \frac{1}{M^2}\left|I_2\left(\Delta_1\right)
\right|^2$, $Q_2 = G_{\mathrm{cp}} \left(\Delta_2\right)=\frac{1}{M^2}\left|I_1\left(\Delta_2\right)\right|^2 $, $Q_3 = G_{\mathrm{cp}} \left(\Delta_3\right) = \frac{1}{M^2}\left(
I_1^2\left(\Delta_3\right) + I_2^2\left(\Delta_3\right)\right)$, $Q_4 = G_{\mathrm{cp}} \left(2\Delta_1\right)=\frac{1}{M^2}\left|I_2\left(2\Delta_1\right)
\right|^2$.

\subsubsection{SLH}
The grating lobe positions for the two subarrays of co-prime array can be obtained by \cite{10465094},
\begin{equation}
	\begin{aligned}
		& \mathcal{S}_1=\frac{2n}{M_2},\quad n=\pm1,\pm2,\ldots,\pm \left(M_2-1\right),\\
		& \mathcal{S}_2=\frac{2n}{M_1},\quad n=\pm1,\pm2,\ldots,\pm \left(M_1-1\right).
	\end{aligned}
\end{equation}
The elements of $\mathcal{S}_1$ and $\mathcal{S}_2$ will not overlap due to the co-prime sparsity of two subarrays.
This makes $G_{\mathrm{cp}}\approx I_1^2$ when $\Delta \in \mathcal{S}_1$ and $G_{\mathrm{cp}}\approx I_2^2$ when $\Delta \in\mathcal{S}_2$.
Therefore, the side lobes of $G_{\mathrm{cp}}$ can be approximated by
\begin{theorem}
	\label{liangyingchang}
	The side lobes of co-prime array will occur at
	\begin{equation}
		\mathcal{S} = 	\mathcal{S}_1 \cup 	\mathcal{S}_2.
	\end{equation}
	The $\mathrm{SLH}$ is
	\begin{equation}
		\mathrm{SLH} \approx\left\{\begin{aligned}
			& \frac{4M_1^2}{M^2}, & \Delta\in\mathcal{S}_1,\\
			& \frac{\left(M_2-1\right)^2}{M^2}, & \Delta\in\mathcal{S}_2.\\
		\end{aligned}\right.
	\end{equation}
\end{theorem}
\begin{figure}[htbp]
	\centering
	\includegraphics[width = 6cm]{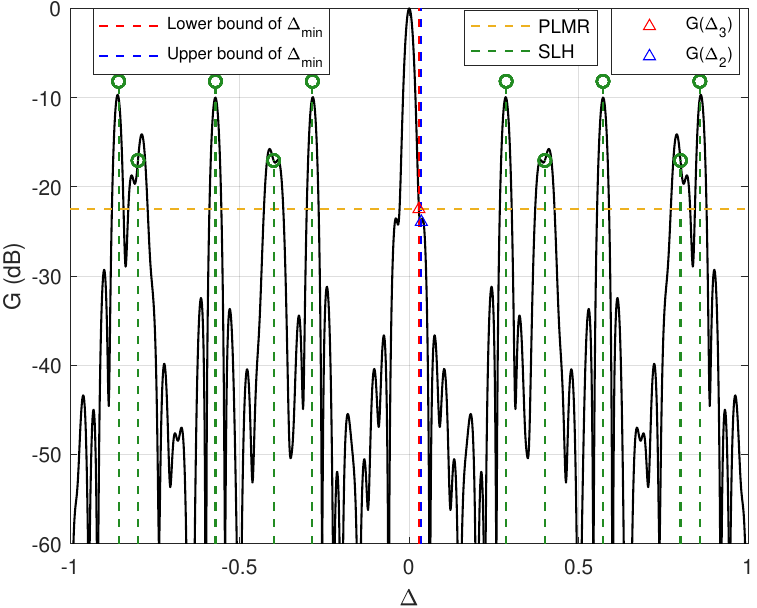}
	\caption{Beam pattern of co-prime array with $\left(M_1,M_2\right)=\left(5,7\right)$.}
	\label{beampattern_cp}
\end{figure}

Fig. \ref{beampattern_cp} validates the effectiveness of the proposed co-prime array beam pattern characteristics.
Firstly, we have $M_{th,1}=M_{th,2}=8$ for $M_1=5$, thereby the subarray parameters satisfy $M_1<M_2\leq M_{th,1}$. 
From Theorem \ref{liulichu}, when $M_2 < M_{th,2} $, co-prime based sparse MIMO forms a narrower main lobe within the main lobe region of subarray 2, and its width is determined by either (\ref{yexiaoying}) or (\ref{wangxi}). 
In this case, the main lobe beam width satisfies $\mathrm{BW}\leq \frac{1}{(M_1-1)M_2}$. It can be observed that $\mathrm{BW}$ is also much smaller than that of compact MIMO with the same number of antennas, making sparse MIMO based on co-prime arrays can also provide higher spatial resolution. 
Lastly, the side lobes of co-prime based sparse MIMO are more complicated than nested based sparse MIMO, and they can be divided to two groups, each originating from one of the two subarrays, as shown in Theorem \ref{liangyingchang}.

\subsection{Sparse MIMO for ILAC}
Based on the above analysis, different sparse architectures result in different beam pattern characters, i.e., $\mathrm{BW}$, $\mathrm{PLMR}$, $\mathrm{SLH}$.
Typically, sparse MIMO based on both nested and co-prime arrays can achieve $\mathrm{BW}$ being $\frac{1}{\mathcal{O}\left(M^2\right)}$, compared to the $\frac{1}{\mathcal{O}\left(M\right)}$ of compact ULA.
Meanwhile, different sparse MIMO configurations also form different virtual arrays, achieving varying levels of localization resolution and DoFs, as illuminated in Section \ref{wuyifan}. 
These properties indicate that the communication and localization performance of sparse MIMO are coupled with respect to the sparse array architectures. 
To demonstrate the joint communication and localization performance gain of sparse MIMO based ILAC system, we define the following two metrics:
\subsubsection{Communication Performance Metric}
The achievable sum data rate $R_s$ is adopted as the communication performance metric, i.e., 
\begin{equation}
	R_s = \sum\nolimits_{k=1}^{K_c}R_k.
\end{equation}

\subsubsection{Localization Performance Metric} The root mean squared error ($\mathrm{RMSE}$) of the estimator is used as an indicator of sensing performance, which is expressed by
\begin{equation}
	\mathrm{RMSE}=\sqrt{\frac{1}{K-K_c}\sum\nolimits_{k=K_c+1}^{K}\left(\theta_k-\hat{\theta}_k\right)^{2}},
\end{equation}
where $\hat{\theta}_k$ is the estimated AoA of L-UE $k$.

\section{Simulation Results} \label{zhangchaoyang}
In this section, numerical results are provided to validate the performance gain of sparse MIMO based ILAC over that based on the conventional compact MIMO.
We consider an uplink multi-user ILAC system with transmit SNR given by $20\ \mathrm{dB}$. We use the ``one-ring'' multi-path channel model with $L_k=10$, $R=5\ \mathrm{m}$, and $r=40\ \mathrm{m}$ denoting the number of multi-path, the radius of each ring, and the range of the center of the ring, respectively. The number of snapshot and Monte Carlo is set to $1000$ and $500$, respectively. Besides, the Bartlett algorithm is adopted to realize localization and the searching grid spacing is set as $0.001\ \mathrm{ rad}$. Unless otherwise stated, we consider the challenging hot-spot scenario where $K=30$ UEs are equally spaced within the interval $[-\theta_{max}, \theta_{max}]$ with $\theta_{max}=6^{\circ}$. 

\begin{figure}[htbp]
	\centering
	\begin{subfigure}{1\linewidth}
		\centering
		\includegraphics[width=6cm]{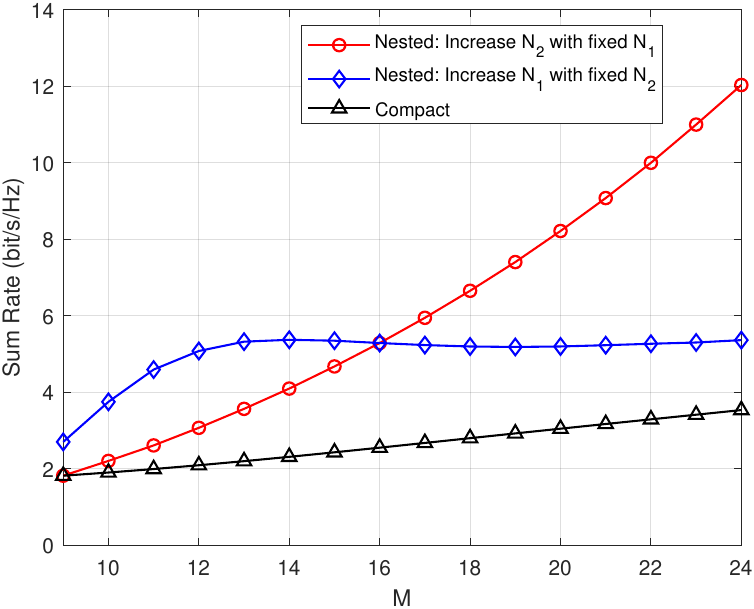}
		\caption{Adjusting different subarrays of nested array.}
		\label{nested1}
	\end{subfigure}
	\centering
	\begin{subfigure}{1\linewidth}
		\centering
		\includegraphics[width=6cm]{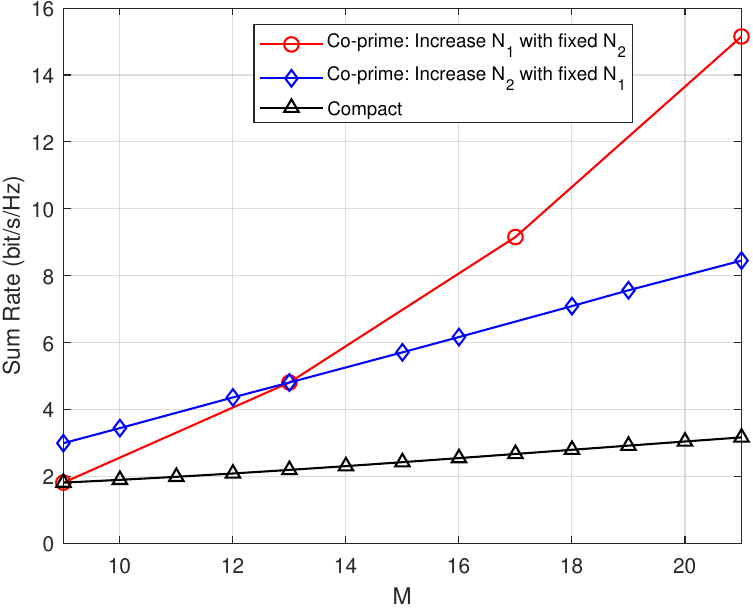}
		\caption{Adjusting different subarrays of co-prime array.}
		\label{coprime_changeN}
	\end{subfigure}
	\caption{The communication performance of different sparse array architectures.}
	\label{huangxuncai}
\end{figure}

\subsection{Sparse MIMO Communication Performance}
Fig. \ref{huangxuncai} illustrates the communication sum rates of C-UEs with different sparse array architectures versus the number of array elements $M$.
For sparse arrays, the increase in $M$ is achieved by fixing the number of antennas in one subarray and adjusting that in another subarray.
On the one hand, it is observed that sparse arrays achieve much higher sum rates than that of compact ULA and the performance gain becomes even more significant as $M$ increases. This is expected since the main lobe beam width of sparse arrays is $\frac{1}{\mathcal{O}\left(M^2\right)}$, compared to $\frac{1}{\mathcal{O}\left(M\right)}$ of compact MIMO, which is shown in Theorem \ref{wuyulin} and \ref{liulichu}.
On the other hand, for the nested array based sparse MIMO, by increasing the inner subarray $N_1$, the performance improvement is not as significant as increasing the outer sub-array $N_2$. In fact, performance does not necessarily increase monotonically with the increase of $M$, since the $\mathrm{PLMR}$ becomes smaller in this case.
Specifically, with $N_1$ increasing from 1 to 16, the two thresholds in Theorem \ref{wuyulin} satisfy $N_{th} \leq 3$ and $N_{ap} \leq 8$, thereby the subarray parameters of nested array always satisfy $N_2 \geq N_{ap}$ when $N_2$ is fixed at 8. 
Then we can get $\mathrm{PLMR}\geq \max\{ \frac{1}{P_{int}}, \frac{1}{P_{2}}\}$ based on Lemma \ref{fengyingqi}. In this case, we analyze $\mathrm{PLMR}$ through $\frac{1}{P_{2}}$ for simplicity because $\frac{1}{P_{int}}$ and $\frac{1}{P_{2}}$ have similar values. 
It can be proved that 
$\mathrm{PLMR}$ decrease monotonically with $N_1$. 
Therefore, when $N_1$ is small, increasing $N_1$ can reduce $\mathrm{BW}$, and effectively suppress some IUI, thus the sum rate increases. However, as $N_1$ continues to increase, $\mathrm{PLMR}$ correspondingly decreases, and the ability of suppressing IUI is greatly weakened, resulting in the sum rate degradation.
By contrast, increasing $N_2$ from 1 to 16 with $N_1$ fixed at 8, the corresponding sum rates increases very quickly, with a large slope. This is because increasing $N_2$ narrows the main lobe and increases the $\mathrm{PLMR}$ simultaneously.
Lastly, as for co-prime array based sparse MIMO, it is shown that enhancing both $M_1$ and $M_2$ will benefit the communication, but the slope of the communication rate is steeper when enhancing $M_1$. This is because increasing both $M_1$ and $M_2$ makes the $\mathrm{BW}$ narrower, and a larger $M_1$ will result in a higher $\mathrm{PLMR}$.

\subsection{Sparse MIMO Satisfying Pareto Condition}
\begin{figure}[H]
	\centering
	\includegraphics[width = 6cm]{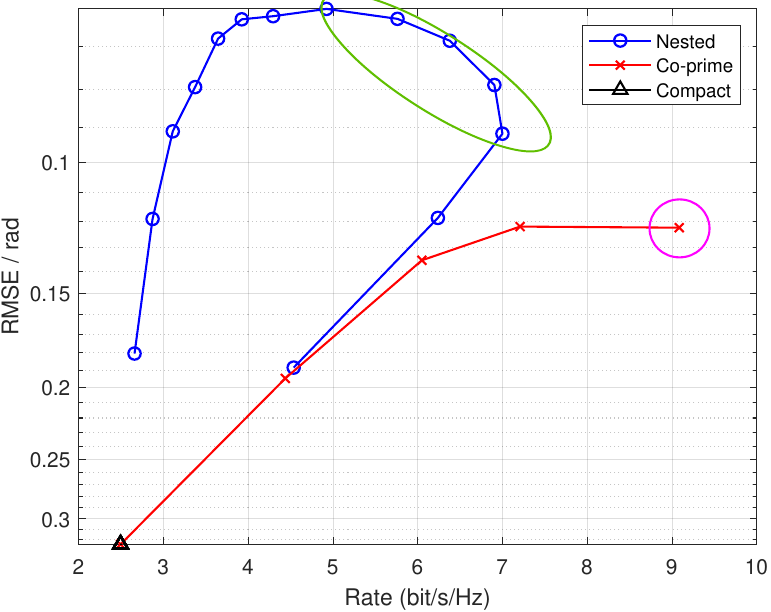}
	\caption{Joint sum rate and RMSE performance when $M=16$.}
	\label{pareto_fixedM}
\end{figure}

Fig. \ref{pareto_fixedM} shows the communication sum rates and localization RMSE achieved by sparse MIMO and traditional compact MIMO. Firstly, the performance of compact ULA only locates at a single point since the ULA has a fixed array structure for a given $M$. In contrast, sparse MIMO can achieve varying communication and localization performances for different subarray configurations.
Secondly, for the same $M$, it is shown that nested array based sparse MIMO can achieve better localization resolution due to a larger virtual aperture, while co-prime array based sparse MIMO can achieve better communication performance due to its larger physical array aperture. Last, we observe a trade-off between communication and localization performance for nested arrays, whereas this trade-off does not exist for co-prime arrays. Therefore, for a given user distribution and $M$, we can obtain sparse MIMO configurations that satisfy the Pareto condition as indicated by the green and pink solid circles.

\subsection{Sparse MIMO Optimization}
For a more general scenario, UEs are not only distributed in hot-spot areas as $\theta_{max}=6^{\circ}$, but may also be distributed at arbitrary angles in the half-space. In this case,  $\theta_{max}$ increases from $6^{\circ}$ to $60^{\circ}$. Here, we set $K_c=22$ and $K_l = 8$. 
To this end, communication and localization performances of sparse MIMO under communication-centric (C-C), localization-centric (L-C) sparse array architectures are provided. Here, C-C signifies that for each given user distribution, the simulation is conducted with the sparse configuration that yields the best communication performance, while L-C denotes the sparse configuration that provides the best positioning performance. 

\begin{figure}[H]
	\centering
	\includegraphics[width = 6cm]{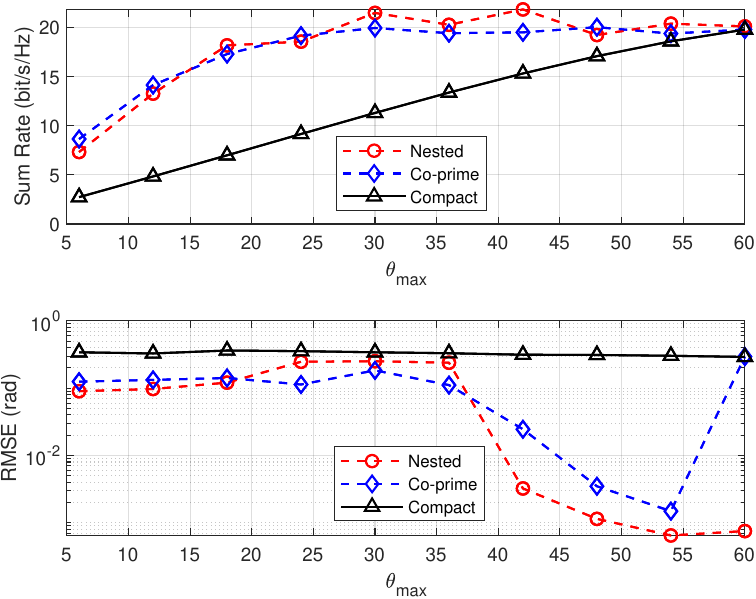}
	\caption{Sum rate and RMSE v.s. $\theta_{max}$ for C-C sparse architectures.}
	\label{rateCC_vsThetaMax}
\end{figure}

Fig. \ref{rateCC_vsThetaMax} shows the sum rates and RMSE of the C-C based sparse array versus different angular range $\theta_{max}$ of users. It can be observed that sparse MIMO can always have better or similar communication sum rates while getting better localization performance compared to its compact counterpart. 
As $\theta_{max}$ increases, the rate gain becomes smaller as the user distribution becomes more dispersed, and the impact of grating lobes of the sparse array will be more severe.
As $\theta_{max}$  increases to $ 42^{\circ}$, the localization advantage of the sparse array begins to emerge. This is because the virtual array technology helps sparse array to localize these L-UEs. In contrast, the compact ULA, constrained by its aperture-limited resolution, consistently fails to achieve effective localization.
Note that co-prime array based sparse MIMO degrades to ULA when $\theta_{max}=60^{\circ}$ due to the C-C configuration.
\begin{figure}[H]
	\centering
	\includegraphics[width = 6cm]{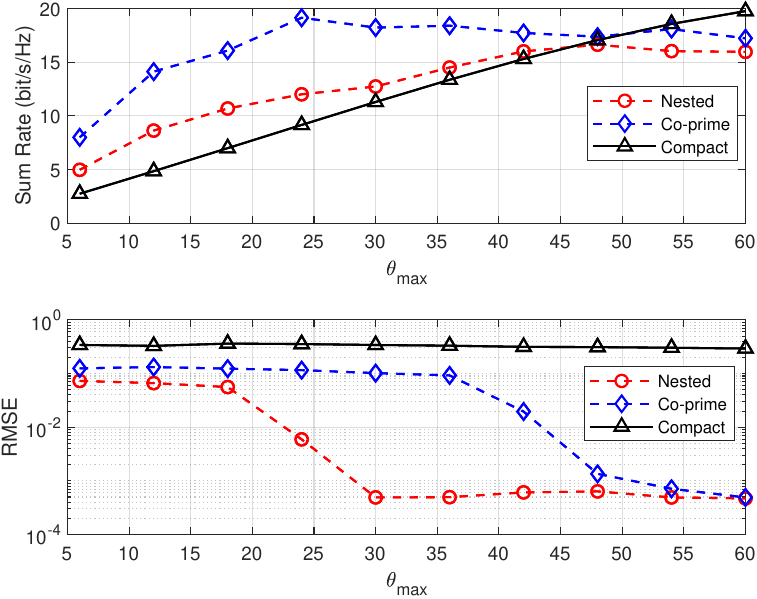}
	\caption{Sum rate and RMSE v.s. $\theta_{max}$ for L-C sparse architectures.}
	\label{rateSC_vsThetaMax}
\end{figure}

Fig. \ref{rateSC_vsThetaMax} illustrates the sum rates and RMSE of the L-C based sparse MIMO versus different angular range $\theta_{max}$ of users.
Firstly, it is shown that nested array based sparse MIMO achieves the best localization performance, as it can locate all L-UEs even when $\theta_{max}=30^{\circ}$, followed by co-prime based sparse MIMO, both of which outperform the conventional compact counterpart. Additionally, it can be observed that under the current conditions, both nested and co-prime based sparse MIMO can finally localize L-UEs. Furthermore, when L-C design is considered, the communication performance of sparse MIMO is limited, especially when $\theta_{max}$ is large, where the performance of sparse MIMO may be worse than conventional ULA. It is also worth noting that although the localization performance of co-prime arrays is not as good as nested arrays, its communication performance is slightly better than nested based sparse MIMO. 

\begin{figure}[H]
	\centering
	\includegraphics[width = 6cm]{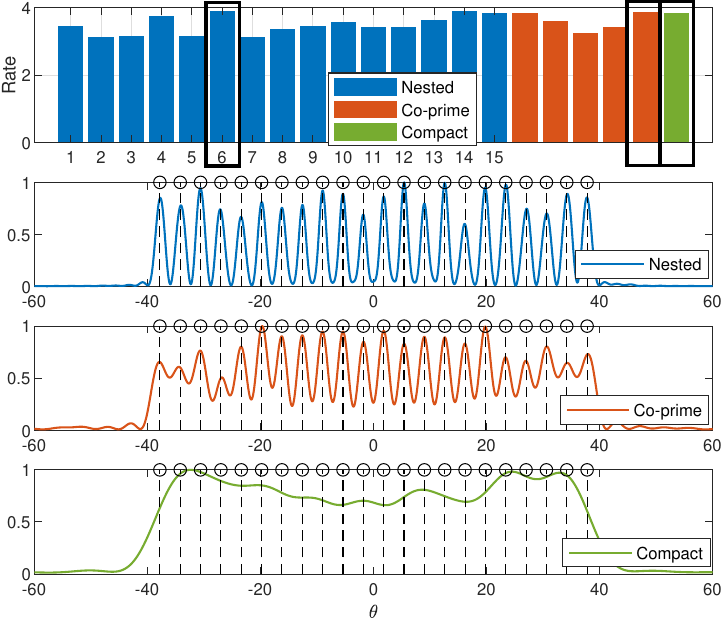}
	\caption{DoF comparison of compact ULA and sparse arrays.}
	\label{TC_design}
\end{figure}

In Fig. \ref{TC_design}, we select sparse array architectures whose communication performances are close to that of a compact ULA and compare their performance in terms of DoF for localization.
Here, we set $K=22$ and $K_c=2$. It can be observed that when the number of L-UEs is greater than the number of physical array antennas $M$, conventional compact MIMO cannot simultaneously localize all the L-UEs, exhibiting significant errors. However, with nested or co-prime array based sparse MIMO, when we select configurations with communication performance similar to that of compact MIMO, the localization DoFs are improved compared to compact MIMO, allowing the localization of more L-UEs than the number of physical array antennas thanks to the virtual array technology.

\section{Conclusion}
In this paper, we consider an ILAC system with a sparse MIMO based on nested and co-prime arrays deployed at the BS.  
The question whether virtual array also benefits communication is well addressed from three aspects, i.e., the destruction of phase information, degradation of SNR, and aggravation of multi-user interference. Therefore, the answer is negative and we further propose the hybrid processing for sparse MIMO based ILAC, i.e., physical array based communication while virtual array based localization. 
Next, we characterize the beam pattern of sparse arrays by three metrics, i.e., $\mathrm{BW}$, $\mathrm{PLMR}$, and $\mathrm{SLH}$, demonstrating that sparse arrays can also bring benefits to communications even though there exists grating lobes due to the narrower main lobe beam width being $\frac{1}{\mathcal{O}\left(M^2\right)}$, compared to $\frac{1}{\mathcal{O}\left(M\right)}$ of compact MIMO.
Extensive simulation results are presented to demonstrate the performance gain of sparse MIMO based ILAC over that based on the conventional compact MIMO.
{\appendices
	

	\section{Proof of (\ref{mahuateng})} \label{wuzhenyang}	
	The expectation in (\ref{baideng}) can be decomposed into the following terms and the time index $n$ is omitted for simplicity,. 
	Firstly, we assume that $x$ follows a complex Gaussian distribution with $E\{x\} = 0$ and $var\{x\} =1$. For $x = x_R + i x_I$ with $x_R,x_I \sim \mathcal{N}\left(0, \frac{1}{2}\right)$, we have $E\{x_R^4 \}= E\{x_I^4\} = 3 \cdot (\frac{1}{2})^2 = \frac{3}{4}$ \cite{nadarajah2005generalized} and $E\{x_R^2 x_I^2\} = E\{x_R^2 \} E\{x_I^2 \} = \frac{1}{4}$. Thus, this leads to $E\{|x|^4\} =E\{(x_R^2 + x_I^2)^2\} = E\{x_R^4 + 2x_R^2 x_I^2 + x_I^4\} = 2$. $E\{|n_i|^4\} = 2\sigma^4$ can be obtained in a similar way, where $n_i$ is the $i$th element of $\mathbf{n}$.
	Next, it can be proved that
			$ E\{| \mathbf{a}^T \mathbf{n}^* \beta\sqrt{P} x^* |^2 \}  
			 = E\{ ( \mathbf{a}^T \mathbf{n}^* \beta\sqrt{P} x^* )( \mathbf{a}^T \mathbf{n}^* \beta\sqrt{P} x^* )^* \} 
			 = MP|\beta|^2\sigma^2$, and $\{| \left(\mathbf{n}^T \mathbf{a}^*\right) \sqrt{P} x |^2\} =MP|\beta|^2\sigma^2$ can be obtained in the same way.
	Meanwhile, we can get 
	\begin{equation}
		\resizebox{0.89\hsize}{!}{$\begin{aligned}
			& E\left\{\left|\frac{\mathbf{a}^H \mathbf{n} \mathbf{n}^H \mathbf{a}}{M}\right|^2\right\} 
			= \frac{1}{M^2}E \left\{\left|\sum\nolimits_{i=1}^{M}\sum\nolimits_{j=1}^{M}a_i^*a_jn_in_j^*\right|^2\right\} \\
			& = \frac{1}{M^2}E\left\{\sum\nolimits_{i=1}^{M}\sum\nolimits_{j=1}^{M}\left|n_in_j^*\right|^2\right\} \\
			& = \frac{1}{M^2}E\left\{\sum\nolimits_{i=1}^{M}\sum\nolimits_{j=i}^{M}|n_i|^4 + \sum\nolimits_{i=1}^{M}\sum\nolimits_{j=1,j\neq i}^{M}\left|n_in_j^*\right|^2\right\} \\
			& = \frac{1}{M^2}\left( 2M\sigma^4 +\left( M^2-M\right)\sigma^4 \right) = \frac{M+1}{M}\sigma^4.
		\end{aligned}$}
	\end{equation}

	Moreover, $E\{ \mathbf{a}^T \mathbf{n}^* \beta^*\sqrt{P} x^*  (\frac{\mathbf{a}^H \mathbf{n} \mathbf{n}^H \mathbf{a}}{M})^* \} = E\{\mathbf{n}^T \mathbf{a}^*\beta\sqrt{P} x (\frac{\mathbf{a}^H \mathbf{n} \mathbf{n}^H \mathbf{a}}{M})^*\}= 0$ always holds since $E\{x\}=E\{x^*\}=0$, and $E\{ \mathbf{a}^T \mathbf{n}^*\beta^* \sqrt{P} x^*  (   \mathbf{n}^T \mathbf{a}^*\beta\sqrt{P} x )^* \}=0$. 
	Therefore, the SNR of the received signal on the virtual array is
	\begin{equation}\label{wujinmin}
		\begin{aligned}
			\mathrm{SNR}_{vir} & = \frac{2 M^2  P^2|\beta|^4 }
			{ 2MP|\beta|^2\sigma^2 + \frac{M+1}{M}\sigma^4}.
		\end{aligned}
	\end{equation}
	From (\ref{wujinmin}), we can easily infer $\mathrm{SNR}_{vir} \leq \mathrm{SNR}_{phy}$. As transmit $\mathrm{SNR}$ keeps increasing, the $\mathrm{SNR}_{vir}$ approximates to $\mathrm{SNR}_{phy}$. The proof of (\ref{mahuateng}) is thus completed.

	\section{Proof of Theorem \ref{liulichu}} \label{lvjiangbin}
	In the following, we analyze the $\mathrm{FLMP}$ $\Delta_{min}$ of $G_{\mathrm{cp}}$ in different value intervals of $\Delta$. 
	Besides, the $\mathrm{BW}$ of $\left|I_1\right|$ is $\mathrm{BW}_1=2\Delta_1=\frac{2}{M_1M_2}$ and the height of the main lobe is $2M_1$. 
	Coherently, the $\mathrm{BW}$ of $\left|I_2\right|$ is $\mathrm{BW}_2=2\Delta_2=\frac{4}{M_1\left(M_2-1\right)}$ and the height of the main lobe is $M_2-1$. 
	\subsubsection{$0\leq \Delta\leq \Delta_1$} It is easy to find that $I_1$, $I_2$, $\cos \Phi$ are positive and monotonously decrease with $\Delta$.  Based on (\ref{liuyuchen}), we get $G_{\mathrm{cp}}$ monotonously decrease with $\Delta$, thus $\Delta_{min}\geq \Delta_1$.
	
	\subsubsection{$\Delta_1 \leq \Delta\leq \Delta_3$} Both $I_1$, $I_2$, $\cos \Phi$ are highly nonlinear, thus it is difficult to derive a closed-form of $\Delta_{min}$. However, qualitative analysis and numerical calculations can provide us a useful boundary. If we want to know whether $\mathrm{FLMP}$ exists in this interval, it is necessary to judge whether $\frac{\mathrm{d}G_{\mathrm{cp}}}{\mathrm{d}\Delta}$ exists positive value, where $\frac{\mathrm{d}G_{\mathrm{cp}}}{\mathrm{d}\Delta}$ is derived as
	\begin{equation}\label{wangxingwei}
		\begin{aligned}
			\frac{\mathrm{d}G_{\mathrm{cp}}}{\mathrm{d}\Delta}
			& = \frac{2}{M^2}\left(
			I_1\frac{\mathrm{d}I_1}{\mathrm{d}\Delta} + I_2\frac{\mathrm{d}I_2}{\mathrm{d}\Delta} + \frac{\mathrm{d}I_1}{\mathrm{d}\Delta}I_2\cos\Phi + \right. \\
			& \left. I_1\frac{\mathrm{d}I_2}{\mathrm{d}\Delta}\cos\Phi 
			+ I_1I_2\frac{\mathrm{d}\cos\Phi}{\mathrm{d}\Delta}
			\right).
		\end{aligned}
	\end{equation}
	Utilizing the aforementioned $\mathrm{FNP}$ properties, we can easily infer $I_1<0, I_2>0, \cos \Phi>0$, $\frac{\mathrm{d}I_1}{\mathrm{d}\Delta}$, $\frac{\mathrm{d}I_2}{\mathrm{d}\Delta}$ and $\frac{\mathrm{d}\cos\Phi}{\mathrm{d}\Delta}<0$. 
	For $\Delta\rightarrow \Delta_3$, we have $\cos \Phi\rightarrow 0$ and $ \frac{\mathrm{d}\cos\Phi}{\mathrm{d}\Delta}\rightarrow -1$ , yielding
	\begin{equation}\label{liuwenzhi}
		\begin{aligned}
			\left. \frac{\mathrm{d}G_{\mathrm{cp}}}{\mathrm{d}\Delta} \right|_{\Delta\rightarrow \Delta_3}
			& \approx \frac{2}{M^2}\left(
			I_1\frac{\mathrm{d}I_1}{\mathrm{d}\Delta} + I_2\frac{\mathrm{d}I_2}{\mathrm{d}\Delta} + \left(- I_1I_2\right)
			\right),
		\end{aligned}
	\end{equation}
	where $I_1\frac{\mathrm{d}I_1}{\mathrm{d}\Delta}$ and $-I_1I_2$ are positive, while $I_2\frac{\mathrm{d}I_2}{\mathrm{d}\Delta}$ is negative.
	Based on (\ref{liuwenzhi}), it can be qualitatively proved that $\exists M_1, M_2, \ \frac{\mathrm{d}G_{\mathrm{cp}}}{\mathrm{d}\Delta}\geq 0$ holds.
	The reasons are provided as follows: Firstly, we approximate the slopes of $I_1$ and $I_2$ as $\frac{\mathrm{d}I_1}{\mathrm{d}\Delta}\approx -\frac{2M_1}{\Delta_1}$ and $\frac{\mathrm{d}I_2}{\mathrm{d}\Delta} \approx -\frac{M_2-1}{\Delta_2}$, respectively. Then we can construct the inequality equation $\frac{\mathrm{d}I_2}{\mathrm{d}\Delta}>\frac{\mathrm{d}I_1}{\mathrm{d}\Delta}$, yielding $\frac{\left(M_2-1\right)^2}{M_2}<4M_1$. This indicates that if $M_2$ is small enough, i.e. $M_2\leq M_{th,1}$, the above inequality can be true, and make $I_1\frac{\mathrm{d}I_1}{\mathrm{d}\Delta} + I_2\frac{\mathrm{d}I_2}{\mathrm{d}\Delta}>0$. To this end, by combining with $\left. \frac{\mathrm{d}G_{\mathrm{cp}}}{\mathrm{d}\Delta}\geq\right|_{\Delta\rightarrow \Delta_1} < 0$, it can be obtained that a local minimum point exists in the current interval. However, $M_{th,1}$ is depend on the value of $M_1$ and difficult to get a closed-form expression, thus a numerical threshold exhaustive search is conducted on MATLAB to find $M_{th,1}$ for given $M_1$. Finally, we can get 
	\begin{equation}
		\left\{
		\begin{aligned}
			&\Delta_1 <\Delta_{min} < \Delta_3, &M_2 \leq M_{th,1},\\
			&\Delta_{min} > \Delta_3, &M_2>M_{th,1}.\\
		\end{aligned}\right.
	\end{equation}
	
	\subsubsection{$\Delta_3 \leq \Delta\leq \Delta_a$} Here, $\Delta_a \approx \frac{3}{2}\Delta_1=\frac{3}{2M_1M_2}$ is a approximation of the $\mathrm{FLMP}$ of $I_1$. 
	Based on the last case, $\left. \frac{\mathrm{d}G_{\mathrm{cp}}}{\mathrm{d}\Delta}\right|_{\Delta\rightarrow \Delta_3} \leq 0$ for $M_2>M_{th,1}$. 
	However, whether there exists $\frac{\mathrm{d}G_{\mathrm{cp}}}{\mathrm{d}\Delta}>0$ in the interval $\Delta_3 \leq \Delta\leq \Delta_a$ is difficult to be analyzed based on (\ref{wangxingwei}), thus vector method based on (\ref{yangyuhang}) is adopted \cite{10404673}. Based on numerically search, it can be verified that $\exists\ M_1, M_2 \text{ and } M_{th,1}<M_2<M_{th,2}, \frac{\mathrm{d}G_{\mathrm{cp}}}{\mathrm{d}\Delta}>0$ holds. For simplicity, the contour line of $\Delta_3$ intersects with the trajectory of $\vec{I}$, which means that $\left|\vec{I}\right|$ exists an local minimum point for $\Delta_3\leq\Delta\leq \Delta_a$. Therefore, the following property holds,
	\begin{equation}
		\left\{
		\begin{aligned}
			&\Delta_3 <\Delta_{min} < \Delta_a, &M_{th,1} < M_2 \leq M_{th,2},\\
			&\Delta_{min} > \Delta_a, &M_2>M_{th,2}.\\
		\end{aligned}\right.
	\end{equation}
	
	\subsubsection{$2\Delta_1 \leq \Delta\leq \Delta_2$} It is easy to get $\left. \frac{\mathrm{d}G_{\mathrm{cp}}}{\mathrm{d}\Delta}\right|_{\Delta=2\Delta_1}<0$ holds since $I_1=0$. Meanwhile, $\left. \frac{\mathrm{d}G_{\mathrm{cp}}}{\mathrm{d}\Delta}\right|_{\Delta=\Delta_2}>0$ since $I_2=0$. This indicates that there exists at least one local minimum point between $2\Delta_1$ and $\Delta_2$ for $G_{\mathrm{cp}}$. Therefore, we have 
	\begin{equation}
		\begin{aligned}
			& 2\Delta_1 < \Delta_{min} < \Delta_2, & M_{th,2} < M_2.\\
		\end{aligned}
	\end{equation}
	Note that for $M_1\geq 14$, $M_{th, 1}$ will not exist, and the case 2) will not exist. 
	This thus completes the proof of Theorem \ref{liulichu}.
	
}

\bibliographystyle{IEEEtran}
\bibliography{reference}
	
\end{document}